\documentclass[journal]{IEEEtran}

\ifCLASSINFOpdf
\else
\fi

\usepackage{algorithmic}
\usepackage{graphicx}
\usepackage{listings}
\usepackage{xspace}
\usepackage[table]{xcolor}
\usepackage{multirow}
\usepackage{xspace}
\usepackage{subfigure}
\usepackage{listings}
\usepackage{textcomp}
\usepackage{stfloats}
\usepackage[ruled,linesnumbered,noend]{algorithm2e}
\usepackage{booktabs}
\usepackage{url}
\usepackage{hyperref}
\usepackage{enumitem}
\usepackage[cmex10]{amsmath}
\usepackage{amssymb}

\newcommand{\mysystem}{\mbox{\texttt{AdvSQLi}}\xspace}
\newcommand{\mysystemr}{\mbox{\texttt{AdvSQLi(R)}}\xspace}
\newcommand{\mysystemm}{\mbox{\texttt{AdvSQLi}}\xspace}
\newcommand{\mole}{\mbox{\texttt{WAF-A-MoLE}}\xspace}
\newcommand{\rl}{\mbox{\texttt{DRL}}\xspace}
\newcommand{\mysystemv}{\mbox{\texttt{AdvSQLi(A)}}\xspace}
\newcommand{\local}{\mbox{\text{SQLi detection}}\xspace}

\newcommand{\etal}{\textit{et al.}}
\newcommand{\ie}{\textit{i.e.}}
\newcommand{\eg}{\textit{e.g.}}
\newcommand{\etc}{\textit{etc}}

\newcommand\vartextvisiblespace[1][.5em]{%
    \makebox[#1]{%
    \kern.07em
    \vrule height.3ex
    \hrulefill
    \vrule height.3ex
    \kern.07em
  }
}

\newcommand{\tabincell}[2]{\begin{tabular}{@{}#1@{}}#2\end{tabular}}

\DeclareRobustCommand{\perthousand}{%
  \ifmmode
    \text{\textperthousand}%
  \else
    \textperthousand
  \fi}

\lstdefinestyle{customstyle}{
  breaklines=true,
  frame=single,
  basicstyle=\linespread{1.15}\small,
  numbers=left,
  numberstyle=\small,
  columns=fullflexible,
  showlines=true,
  mathescape=true,
  escapechar={^},
  linewidth=1.04\columnwidth,
  xleftmargin=0.01\columnwidth
}

\begin{document}
%
\title{\mysystem: Generating Adversarial SQL Injections against Real-world WAF-as-a-service}

\author{Zhenqing~Qu\textsuperscript{*},
        Xiang~Ling\textsuperscript{*†},
        Ting Wang,
        Xiang~Chen,~\IEEEmembership{Member,~IEEE,}
        Shouling~Ji,~\IEEEmembership{Member,~IEEE,}
        Chunming~Wu\textsuperscript{†}
\thanks{This work is supported by the National Key R\&D Program of China (No.2022YFB2901305), the ``Pioneer'' and ``Leading Goose'' R\&D Program of Zhejiang (No.2022C01085), and the Fundamental Research Funds for the Central Universities (Zhejiang University NGICS Platform). Xiang Ling is also supported by the National Natural Science Foundation of China under No.62202457.}
\thanks{Zhenqing Qu, Xiang Chen, Shouling Ji, Chunming Wu are with the College of Computer Science and Technology, Zhejiang University, Hangzhou 310027, China (e-mail: quzhenqing@zju.edu.cn, wasdnsxchen@gmail.com, sji@zju.edu.cn, wuchunming@zju.edu.cn). Xiang Ling is with the Institute of Software, Chinese Academy of Science, Beijing 100190, China (e-mail: lingxiang@iscas.ac.cn). Ting Wang is with the Department of Computer Science at Stony Brook University, Stony Brook, NY 11794-2424, United States (e-mail: twang@cs.stonybrook.edu). }
\thanks{\textsuperscript{*}Zhenqing Qu and Xiang Ling contribute equally to this research.}
\thanks{\textsuperscript{†}Xiang Ling and Chunming Wu are the co-corresponding authors.}
}

\markboth{IEEE TRANSACTIONS ON INFORMATION FORENSICS AND SECURITY}%
{Zhenqing Qu \MakeLowercase{\textit{et al.}}: \mysystem: Generating Adversarial SQL Injections against Real-world WAF-as-a-service}

\maketitle

\begin{abstract}
As the first defensive layer that attacks would hit, the web application firewall (WAF) plays an indispensable role in defending against malicious web attacks like SQL injection (SQLi).
With the development of cloud computing, WAF-as-a-service, as one kind of Security-as-a-service, has been proposed to facilitate the deployment, configuration, and update of WAFs in the cloud.
Despite its tremendous popularity, the security vulnerabilities of WAF-as-a-service are still largely unknown, which is highly concerning given its massive usage.
In this paper, we propose a general and extendable attack framework, namely \mysystem, in which a minimal series of transformations are performed on the hierarchical tree representation of the original SQLi payload, such that the generated SQLi payloads can not only bypass WAF-as-a-service under black-box settings but also keep the same functionality and maliciousness as the original payload.
With \mysystem, we make it feasible to inspect and understand the security vulnerabilities of WAFs automatically, helping vendors make products more secure.

To evaluate the attack effectiveness and efficiency of \mysystem, we first employ two public datasets to generate adversarial SQLi payloads, leading to a maximum attack success rate of 100\% against state-of-the-art ML-based SQLi detectors.
Furthermore, to demonstrate the immediate security threats caused by \mysystem, we evaluate the attack effectiveness against 7 WAF-as-a-service solutions from mainstream vendors and find all of them are vulnerable to \mysystem.
For instance, \mysystem achieves an attack success rate of over 79\% against the F5 WAF.
Through in-depth analysis of the evaluation results, we further condense out several general yet severe flaws of these vendors that cannot be easily patched.
\end{abstract}


\IEEEpeerreviewmaketitle

\section{Introduction}
With the continuous evolution and global deployment of the internet, web services are playing increasingly important roles of social infrastructure in daily lives. On the other hand, they are also being exposed to world-wide threats from different locations, with different scales and through different methods.
Common web threats~\cite{cloudflare_web_security} include SQL injection (SQLi), cross-site scripting, cross-site request forgery, and distributed denial-of-service, to name just a few.
According to~\cite{pramod2015sqli_sqli_imp_1,zhang2019art4sqli_sqli_imp_2,tian2010research_sqli_imp_3}, SQLi is one of the most common and threatening attack methods, by which the attacker exploits security vulnerabilities by performing SQL queries on the database, thereby directly accessing unauthorized information, creating new user permissions, or even taking control of the underlying system.

\begin{figure*}[htb!]
\centering
\subfigure[The innocent user Alice can get her own information.]{
\label{subfigure:pipeline-1}
\includegraphics[width=0.9\textwidth]{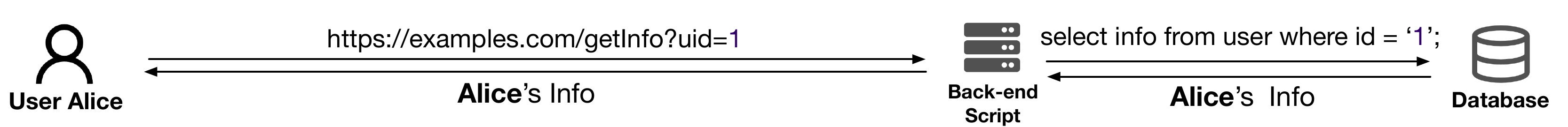}
}
\vspace{-2mm}
\subfigure[The ordinary attacker performs SQLi attack and steals all users' information.]{
\label{subfigure:pipeline-2}
\includegraphics[width=0.9\textwidth]{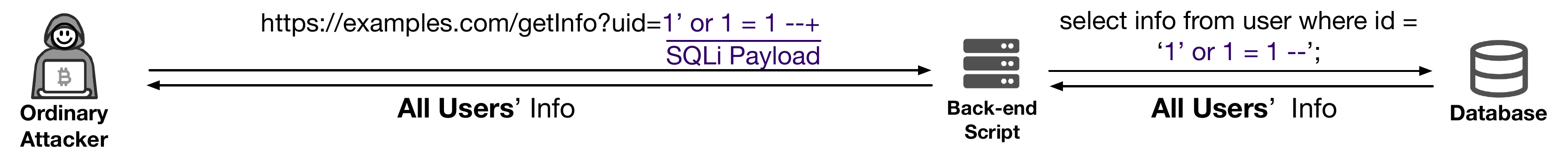}
}
\vspace{-2mm}
\subfigure[Alice can still get her own information.]{
\label{subfigure:pipeline-3}
\includegraphics[width=0.9\textwidth]{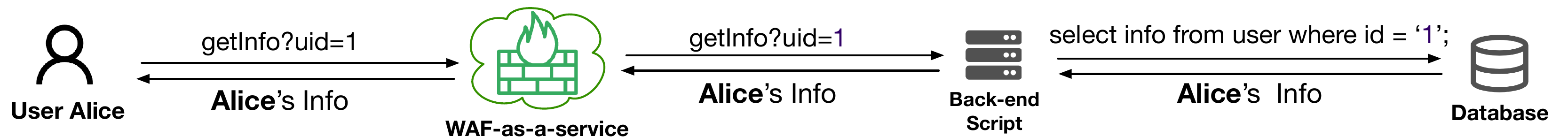}
}
\vspace{-2mm}
\subfigure[The ordinary attacker's request is blocked by the WAF-as-a-service.]{
\label{subfigure:pipeline-4}
\includegraphics[width=0.9\textwidth]{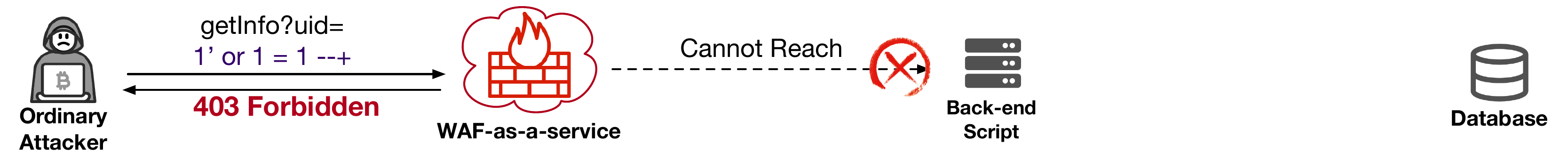}
}
\vspace{-2mm}
\subfigure[\mysystem successfully bypasses the WAF-as-a-service and steals all users' information.]{
\label{subfigure:pipeline-5}
\includegraphics[width=0.9\textwidth]{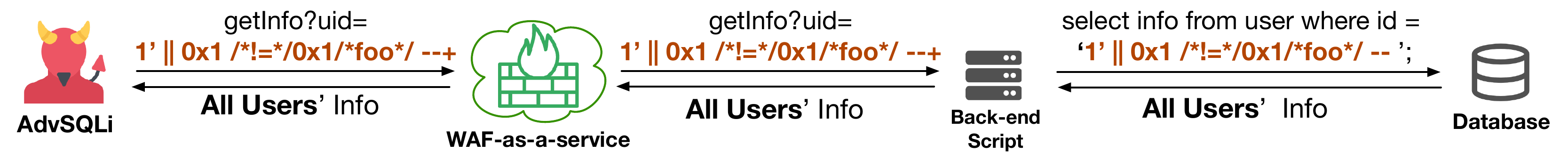}
}
\caption{Illustration of how our proposed attack \mysystem can bypass the WAF-as-a-service to acquire all users' information, compared with the ordinary attacker and the innocent user. Note that the architecture of the infrastructures in production environments may be more complicated than in this diagram, yet it does not impact the effectiveness of \mysystem.}
\label{figure:pipeline}
\end{figure*}

In particular, as illustrated in Figure~\ref{subfigure:pipeline-1}, suppose there is a web server with a back-end script and a corresponding database.
A normal client Alice can send a request of ``getinfo?uid=1'', and receive her information.
However, if an attacker maliciously crafts an SQLi payload (\ie, ``\textbf{1' or 1 = 1 {-\,-\,+}}'') within the request as shown in Figure~\ref{subfigure:pipeline-2}, the attacker can receive all users' information because the ``\textbf{or 1 = 1}'' makes the querying condition to be satisfied.

To mitigate malicious requests like the SQLi in the wild, the web application firewall (WAF) is one of the most widely used and effective defensive systems~\cite{prandl2015study,vartouni2019leveraging,appelt2015behind}.
WAF is normally deployed in the front of the back-end script, indicating that any request from clients must be detected and filtered by the WAF.
Only requests detected as benign will be forwarded to the back-end script for further processing.
Therefore, WAFs play a pivotal role in the immediate response to emerging security vulnerabilities.
For example, during the outbreak of the log4shell vulnerability, Cloudflare, among other vendors, acted promptly by implementing detection rules within their WAF infrastructure~\cite{bypass_log4j}.
To date, extensive research efforts have proposed tremendous WAF strategies to defend against SQLi.
These strategies can be broadly categorized into signature-based (a.k.a rule-based) SQLi detection and machine learning (ML)-based SQLi detection~\cite{applebaum2021signature_short_survey}.
Signature-based SQLi detection~\cite{ristic2010modsecurity,buehrer2005using,ezumalai2009combinatorial} is a traditional but effective strategy which takes advantage of various rules pre-defined by domain experts to detect malicious requests.
Moreover, motivated by the great success of ML obtained from various tasks, a variety of ML-based SQLi detectors~\cite{liu2019locate, corona2009hmm, kar2016sqligot} have been proposed and implemented, which learn to discriminate malicious requests based on the supervision of previously labelled datasets.

Nevertheless, regardless of which kind of strategies are adopted in WAFs, it is quite troublesome to deploy, configure, and update the signatures or models for effectively and timely protection of the web server~\cite{junior2021new}.
With the development of cloud computing, WAF-as-a-service, as one kind of Security-as-a-service, has been proposed to facilitate the deployment, configuration, and update of WAFs.
The administrator who manages the web server only needs to redirect all requests to the interfaces of WAF-as-a-service provided by vendors, such that malicious requests like SQLi can be detected and filtered before being forwarded to the web server.

Despite the tremendous popularity and impressive performance of WAF-as-a-service in detecting malicious requests, the security vulnerabilities of WAFs are still largely unknown, which is highly concerning given the massive usage of WAF-as-a-service in the cloud.
For instance, Peter M managed to bypass Akamai's WAF, triggering a server-side template injection (SSTI) vulnerability in a Spring Boot application, which then led to remote command execution~\cite{bypass_peter}.
In a similar vein, the AON team used encoded payloads to successfully bypass Cloudflare's WAF, enabling them to exploit the Object Graphic Navigation Language (OGNL) injection vulnerability~\cite{bypass_aon}.
To inspect and understand the security vulnerabilities of WAFs automatically, we propose \mysystem, a general and extendable attack framework that can effectively and efficiently generate adversarial SQLi payloads to bypass real-world WAF-as-a-service under the black-box settings, as the feedback output of WAF-as-a-service to the attacker typically is binary.
With a large and diverse set of payloads, we are capable of systematically assessing and understanding the security vulnerabilities within WAFs.

In particular, \mysystem first represents the original SQLi payload with a hierarchical tree and then employs a weighted mutation strategy based on the context-free grammar to generate a set of equivalent SQLi payloads, which keep the same functionality and maliciousness as the original SQLi payload.
Finally, \mysystem exploits the Monte-Carlo tree search as a novel approach to efficiently guide the exploration of adversarial SQLi payloads in the vast space. 

Extensive evaluation results demonstrate that \mysystem outperforms all baseline attack methods with regard to the attack effectiveness and efficiency, achieving a maximum success rate of 100\% against ML-based SQLi detection models within fewer queries.
Furthermore, we also evaluate the attack effectiveness of \mysystem against 7 WAF-as-a-service solutions in the wild from mainstream vendors (\ie, AWS, Cloudflare, F5, Fortinet, Wallarm, CSC, and ModSecurity) under 4 practical request methods, and draw a conclusion that the generated payloads can bypass most of them, indicating the immediate threats caused by \mysystem.
For instance, \mysystem achieves the attack success rate of over 79\% against the F5 WAF.
To summarize, we make the following contributions.
\vspace{-1mm}
\begin{itemize}[leftmargin=*]
\item We present and implement \mysystem, in which a weighted mutation strategy based on context-free grammar is first proposed to generate a vast space of equivalent SQLi payloads.
Then, the Monte-Carlo tree search is exploited to efficiently select the adversarial payload from the vast space.
Manual changes are unnecessary when attacking different WAF-as-a-service concerning unique payloads under kinds of request methods, namely, \mysystem is a broad-spectrum attack framework that works out of the box with strong generalizability and extensibility.
\item To the best of our knowledge, this is the first work systematically assesses the security vulnerabilities of real-world WAF-as-a-service solutions based on adversarial SQLi payloads under strict black-box settings.
\item Extensive evaluation results demonstrate that \mysystem not only achieves a maximum success rate of 100\% against state-of-the-art SQLi detection models but also bypasses 7 mainstream commercial WAF-as-a-service products with high attack success rates. 
\end{itemize}

\vspace{-1mm}
\section{Problem Formulation}
\label{sec:2}
In this section, we present the threat model of this paper.
\subsection{Adversary's Goal}
Given an SQLi detector ${f:\mathcal{X} \to \mathcal{Y}}$, which maps a payload ${\mathbf{x} \in \mathcal{X}}$ to a corresponding label ${y \in \mathcal{Y} = \{0, 1\}}$ (\ie, 0 indicates benign and 1 indicates malicious), an adversary aims to generate an adversarial payload ${\mathbf{x}_{adv}}$ which can be misclassified by ${f}$, \ie, ${f(\mathbf{x}_{adv}) \neq f(\mathbf{x})}$.

We consider a realistic scenario that an SQLi payload ${\mathbf{x}}$ which is originally correctly classified as malicious by $f$ (\ie, ${f(\mathbf{x}) = 1}$), while the adversary attempts to generate an adversarial SQLi payload ${\mathbf{x}_{adv}}$ from $\mathbf{x}$ with minimal efforts, such that ${\mathbf{x}_{adv}}$ can not only bypass the SQLi detector (\ie, ${f(\mathbf{x}_{adv}) = 0}$) but also preserves the same semantics (\ie, functionality and maliciousness) as ${\mathbf{x}}$ as follows.

\begin{equation}
\begin{split}
	\mathop{\arg\min}_{\mathbb{T}}  \, & f(\mathbf{x}_{adv}) \\
    	\text{s.t} \quad
    	 & f(\mathbf{x}) =1 \\
    	 & \mathbf{x}_{adv}  = \mathbb{T} (\mathbf{x}) \\
    	 & Sem(\mathbf{x}_{adv}) = Sem(\mathbf{x})
\end{split}
\label{equation:threat-model}
\end{equation}
in which ${\mathbb{T}}$ denotes the adversarial function that generates adversarial payload ${\mathbf{x}_{adv}}$ from the input payload ${\mathbf{x}}$, while ${Sem(x)}$ is the semantic analyzer to check if ${x}$ and ${{x}_{adv}}$ are semantic-equivalent.

\subsection{Adversary's Capability}
In this paper, we consider a classic but strict black-box attack in the problem space~\cite{pierazzi2020intriguing_problem_stmt,biggio2018wild_knowledge}, in which the adversary limits the attack surface to the testing phase while does not have any information on the target SQLi detector (\eg, architectures, parameters, feature representations, \etc.) except for the input and output.
According to the output information of different SQLi detectors in the wild, the black-box attacks can be further divided into two cases.
The first is the black-box attacks with probability (\textbf{BB w/ prob.}), in which the target SQLi detector (\eg, ML-based SQLi detectors) can output the predicted label (\ie, $y=0 \text{ or } 1$) as well as the corresponding probability (\ie, $p_y \in [0, 1]$).
The second is the black-box attacks without probability (\textbf{BB w/o prob.}), in which the target SQLi detector can only output the predicted label (\ie, $y=0 \text{ or } 1$).
In fact, for the adversary that targets most WAF-as-a-service in the wild, it is a strict black-box attack, \ie, \textbf{BB w/o prob.}, by which the adversary can only judge by the HTTP status code.

\section{Related Work and Challenges}
\label{sec:3}
This section provides a brief study description of untargeted black-box testing and adversarial attacks in the traditional domains, and their difference with our task.

\textbf{Automated Black-box Testing (Untargeted Attacks).}
In the last decade, there has been abundant research on employing algorithms to facilitate vulnerability discovery~\cite{felderer2016security}, including finding vulnerabilities in applications and firewalls to reduce the manual workload.
Tripp \etal~\cite{tripp2013finding} proposed XSS Analyzer, a learning-based black-box testing method for web applications.
It learns from failed attempts to prune the search space.
Further, Appelt~\etal~\cite{appelt2015behind,appelt2018machine} proposed ML-Driven, a search-based approach that combines ML and evolutionary algorithms (EAs) to test the detection capabilities of WAFs.
ML-Driven selects payloads with high bypassing probabilities and mutates them with the help of EAs to generate payloads.
Zhang \etal~\cite{zhang2019art4sqli_sqli_imp_2} proposed ART4SQLi, an adaptive random testing method for SQLi vulnerability detection, which is based on the context-free grammar to parse and extract payload features.
ART4SQLi chooses payloads far from the initial payload from existing collections until it bypasses the target WAF.
Amouei \etal~\cite{amouei2021rat} proposed RAT, which clusters similar payloads based on n-gram features.

\textbf{Automated WAF Bypassing (Targeted Attacks).}
The above work proposes valuable concepts, \eg, learning-based methods that can speed up the vulnerability discovery process.
However, they do not need to preserve the functionality of the payload, but only need to find an arbitrary payload that can trigger the vulnerability.
On the contrary, with the process of finding semantic-preserving payloads for a failed payload, the targeted attacks, can systematically assess the security vulnerabilities based on a diverse set of malicious payloads.
Demetrio \etal~\cite{demetrio2020waf_wafamole} proposed WAF-A-MoLE, a tool that can evade ML-based WAFs for failed payloads.
They first defined 7 string-based SQLi payload mutation operators and employed a priority queue to guide the mutation process.
Specifically, WAF-A-MoLE repeatedly mutates the initial payload to generate several payloads and selects the mutated payload with the lowest malicious score as the initial payload for the next round, until it bypasses the target WAF or reaches the termination condition.
Inspired by the work of evading malware detectors, Wang \etal~\cite{wang2020evading_gymwaf} proposed a WAF evasion framework based on RL.
They followed the mutation operators in~\cite{demetrio2020waf_wafamole} and defined a feature vector of three levels of histograms to represent the \textit{State} of the SQLi payload.
After training with Deep Q-learning (DQN) and Proximal Policy Optimization (PPO) algorithms, the RL \textit{Agent} can select an optimal mutation operator according to the current \textit{State}.
Hemmati \etal~\cite{hemmati2021using} improved~\cite{wang2020evading_gymwaf} in several aspects, including expanding the mutation methods, defining \textit{State} based on the FastText, \etc.

However, the above approaches are not satisfactory in some aspects.
Firstly, the mutation methods depend on string replacement based on regular expression (RE), which is not semantic-preserving.
According to the Chomsky hierarchy~\cite{chomsky1956three}, the RE-based rule descriptions (\ie, rule-based grammar) cannot fully cover the program-language-based attack payloads (\eg, SQLi payloads).
That is, the string-based mutation method may miss mutable parts of the payload.
Besides, it is possible to damage the original function of the SQLi payload, which we will prove through subsequent evaluations. 
In addition, the binary output and the practical limitation of probing attempts by WAF-as-a-service further challenge the adaptability of priority queue and RL methods in real-world scenarios, necessitating a more strategic and sparing use of mutation attempts~\cite{review2_das2022practical,review3_wang2016targeted}.
Literature in other security fields~\cite{anderson2018learning_anderson,wu2018enhancing_rlmalware,fang2019evading_rlaccess} also pointed out that approaches based on RL cannot achieve surprising results in such problems due to limitations such as the representation of feature state.

\textbf{Adversarial Attacks.}
In traditional adversarial attack tasks, such as images~\cite{ling2019deepsec}, the adversary performs small disturbances in the continuous space without affecting the presentation form to evaluate the vulnerabilities of image classification models.
However, as to generating adversarial SQLi payloads, the adversary can only perform problem space transformations in the discrete space.
Moreover, the gradient-based optimization methods (\eg, FGSM~\cite{goodfellow2014explaining_fgsm}) in continuous space cannot be applied to guide our attack procedure, as an SQLi payload is a discrete object and the adversary does not have any information of $f$~\cite{review1_schwarzl2023practical}.
Previous work in adversarial NLP presents valuable mutation methods in the discrete space.
For example, Li~\etal~\cite{li2018textbugger} proposed several mutation methods to deceive ML-based text understanding models.
As the meaning of the text is likely to be preserved by human readers after these slight character changes~\cite{rawlinson2007significance}, these mutation methods are acceptable.
However, these works cannot be applied to generate SQLi payloads, as the generated payload has to keep the same functionality and maliciousness as the original one~\cite{pierazzi2020intriguing_problem_stmt}.

\textbf{Key Challenges}.
In conclusion, the principal challenges we encounter in effectively and efficiently generating payloads that bypass detection are twofold:

\textbf{1) Semantic Preservation:} There is a critical absence of a truly semantic-preserving mutation method for SQLi payload generation.
\textbf{2) Optimization in Discrete Space:} The default strategy in the vast and discrete problem space has been to employ random transformations. This approach lacks efficiency due to the immense search space and is further complicated by semantic constraints.

However, extant methods that leverage reinforcement learning (RL) and priority queues fall short in the context of real-world WAF-as-a-service applications. Therefore, it becomes crucial to develop a mutation approach that not only preserves the semantics of SQLi payloads but also an optimization strategy that is tailored for black-box attacks within this problem space.

\section{\mysystem: Adversarial SQLi Generation}
\label{sec:4}
\begin{figure*}[htb!]
    \centering
    \includegraphics[width=0.99\textwidth]{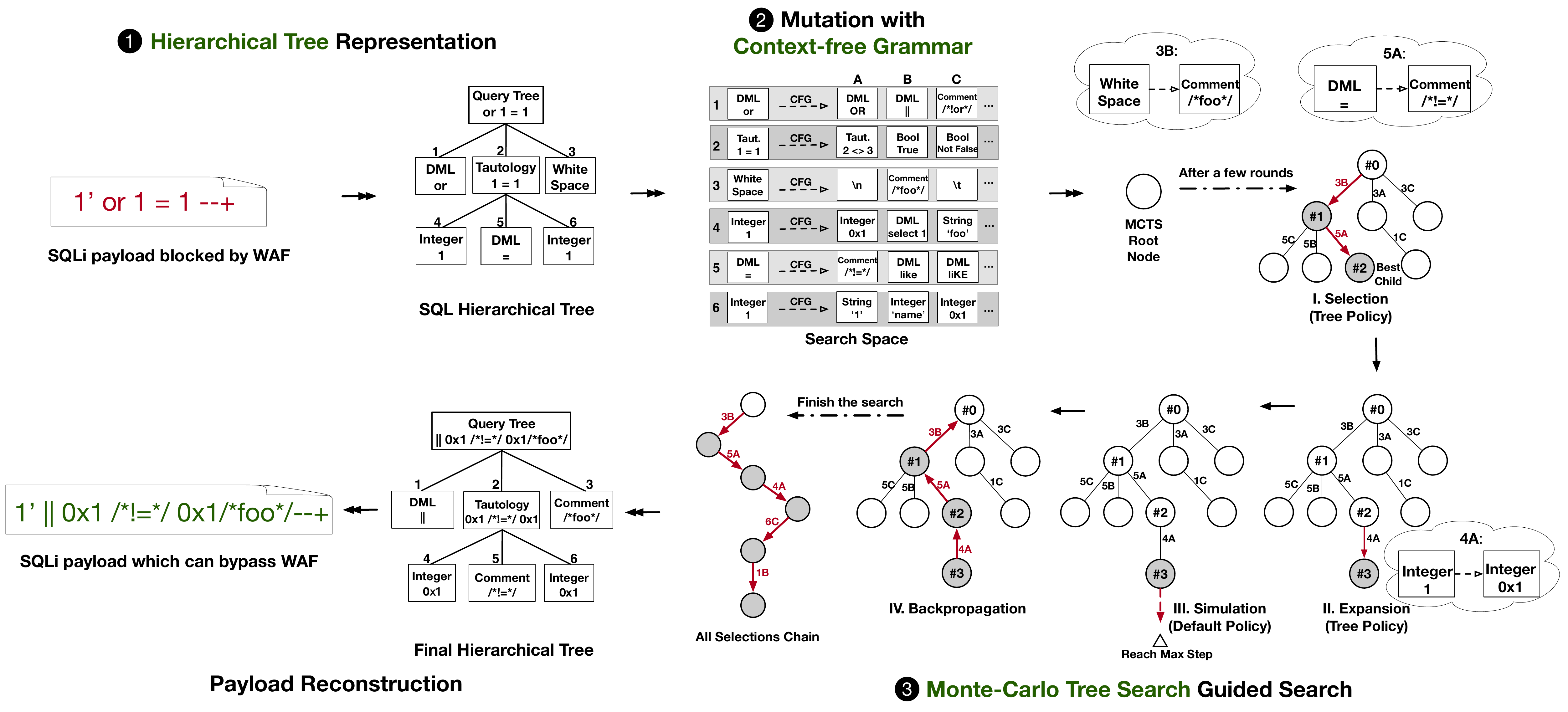}
    \caption{The pipeline of \mysystem: It first represents the original SQLi payload with a hierarchical tree, and then employs a weighted mutation strategy based on the context-free grammar to generate a set of equivalent SQLi payloads, which keep the same functionality and maliciousness as the original one. Then, it exploits MCTS to efficiently guide the exploration of adversarial SQLi payloads in the vast space. It is noted that only parts of the Hierarchical Tree are shown for simplicity.} 
    \label{figure:pipeline-detail}
\end{figure*}

In order to address the aforementioned challenges in bypassing WAF-as-a-service for profits, we propose a general and extendable attack framework \mysystem to generate adversarial SQLi payloads. 
Figure~\ref{figure:pipeline-detail} illustrates the framework overview of \mysystem. 
In particular, \mysystem first represents the original SQLi payload with a hierarchical tree to perform fine-grained and customized processing for each node.
Further, on the basis of the hierarchical tree, \mysystem employs a weighted mutation strategy based on the context-free grammar to generate a set of equivalent SQLi payloads, which keep the same functionality and maliciousness as the original one.
With the two steps, we implement a semantic-preserving mutation method according to the semantics, characteristics, and constraints of the original payload. 
Finally, \mysystem exploits the Monte-Carlo tree search as a novel approach to efficiently guide the exploration of adversarial SQLi payloads in the vast space.
In the Monte-Carlo tree search, the malicious scores feedback by WAFs are unnecessary.
That is, \mysystem is capable of attacking real-world WAFs. 

\subsection{Hierarchical Tree Representation}
\label{sec:4.1}
Normally, a HTTP URL (\eg, \textcolor{blue}{\url{https://examples.com/getInfo?uid=1}}) consists of four elements: schema (\ie, ``{https}''), domain name (\ie, ``{examples.com}''), path to the resource (\ie, ``{getInfo}'') and the parameter(s) (\ie, ``{?uid=1}'') with multiple pairs of key (\ie, ``{uid}'') and value (\ie, ``{1}'', also termed as payload).
In fact, the process of SQL injection attack is limited to the core part of URL, \ie, the parameter value(s), and keeps other parts unchanged. 
For example, the SQLi attack can inject a malicious payload of ``{\textcolor{blue}{\textquotesingle or 1 = 1 -\,-+}}'' at the end of the original payload of ``\textcolor{blue}{1}'', resulting in a malicious URL, \ie, \textcolor{blue}{https://examples.com/getInfo?uid=1\textquotesingle \, or 1 = 1 -\,-+}.

Recall that the goal of \mysystem is to generate adversarial SQLi payloads based on the original SQLi payload.
Therefore, \mysystem should not only be limited to the payload of the original SQLi (\ie, \textcolor{blue}{1\textquotesingle \, or 1 = 1 -\,-+}), but also ensure that the generated adversarial SQLi has the same functionality and maliciousness as the original injection.
To address this problem, according to the different roles played by different parts in the original SQLi, we divide the original SQLi payload into three modules, \ie, left boundary (\ie, ``\textcolor{blue}{{1\textquotesingle}}''), query (\ie, ``\textcolor{blue}{{ or 1 = 1}}'') and right boundary (\ie, ``\textcolor{blue}{{-\,-+}}''), and further limit the attack surface of \mysystem to the module of query in the original SQLi payload. 
In particular, both the left boundary and the right boundary are restricted to unchanged, so that the functionality of the original SQLi is preserved.
The query is the core module of the original SQLi payload and is needed to be further processed and manipulated for generating adversarial SQLi payloads.
What is more, instead of simply treating the query in the original SQLi payload as a sequence of strings~\cite{demetrio2020waf_wafamole,wang2020evading_gymwaf,hemmati2021using}, we further represent it with a hierarchical tree. 
Formally, for the hierarchical tree representation, each parent (non-leaf) node is an SQL statement that assembles all tokens from its ordered child nodes from left to right, and each leaf node is the atomic token (\eg, integer, the keyword of `or', \etc) in SQL.

On the basis of the hierarchical tree representation, we can perform more fine-grained and customized processing for each node according to its unique characteristics and constraints, which could facilitate the generation of adversarial SQLi payloads based on mutation as in the following subsection.

\subsection{Mutation with Context-free Grammar}
\label{sec:cfg}

In order to generate both functionality-preserving and maliciousness-preserving adversarial SQL injection, our proposed \mysystem resorts to the problem space attack methods, \ie, manipulating the original SQL injection payload $\mathbf{x}$ with a minimal cost of transformations, so that the generated adversarial SQL injection payload $\mathbf{x}_{adv}$ can bypass the target SQLi detector (\eg, WAF-as-a-service).
We first limit the transformation to \textit{addition} and \textit{replacement} on the original SQLi payloads and then propose a weighted mutation strategy based on the context-free grammar (CFG) to generate a set of candidate adversarial SQLi payloads, which keeps the same functionality and maliciousness as the original SQLi payload.
That is, for each node and subtree in the tree representation of the original SQLi payload, our weighted mutation strategy can generate a set of candidate actions that can be applied to the original SQLi payload for generating a corresponding set of candidate adversarial SQLi payloads.

\vspace{-1mm}
\begin{table}[htb!]

\setlength\tabcolsep{1pt}
\small
\caption{\upshape Examples of mutation methods. * means that the operator is flexible for different request methods.}

\begin{tabular}{cc}
\toprule
Mutation & Example \\
\midrule
\textit{Case Swapping} & or 1 = 1  $\rightarrow$  oR 1 = 1 \\
\rowcolor{gray!10}\textit{Whitespace Substitution}* & or 1 = 1  $\rightarrow$ \textbackslash{}tor1\textbackslash{}n=1 \\ 
\textit{Comment Injection}* & or 1 = 1  $\rightarrow$ /*foo*/or 1 =/*bar*/1 \\ 
\rowcolor{gray!10}\textit{Comment Rewriting} & /*foo*/or 1 = 1  $\rightarrow$ /*1.png*/or 1 = 1 \\
\textit{Integer Encoding} & or 1 = 1  $\rightarrow$  or 0x1 = 1 \\
\rowcolor{gray!10}\textit{Operator Swapping} & or 1 = 1 $\rightarrow$ or 1 like 1 \\
\textit{Logical Invariant} & or 1 = 1 $\rightarrow$ or 1 = 1 and `a' = `a' \\ 
\rowcolor{gray!10}\textit{Inline Comment} & \begin{tabular}[c]{@{}c@{}} union select $\rightarrow$ /*!union*/ /*!50000select*/\end{tabular} \\ 
\textit{Where Rewriting} & \begin{tabular}[c]{@{}c@{}}where xxx $\rightarrow$ where xxx and True\\ where xxx $\rightarrow$ where (select 0) or xxx\end{tabular} \\ 
\rowcolor{gray!10}\textit{DML Substitution}* & \begin{tabular}[c]{@{}c@{}}or 1 = 1 $\rightarrow$ $||$ 1 = 1\\ and name = `foo' $\rightarrow$ \&\& name = `foo'\end{tabular} \\  
\textit{Tautology Substitution} & \begin{tabular}[c]{@{}c@{}} `1' = `1' $\rightarrow$ 2 $<>$ 3\\ 1 = 1 $\rightarrow$ rand() >= 0 \\
1 = 1 $\rightarrow$ (select ord(’r’) regexp 114) = 0x1\end{tabular} \\
\bottomrule
\end{tabular}
\label{table:examples-of-mutation-methods}
\end{table}

In the weighted mutation strategy, we first present the context-free grammar $G$ which is formally defined as a 4-tuple $G=(S, V, \Sigma, R)$.
In particular, $S$ is the set of starting symbols, which is associated with the nodes and subtrees of the hierarchical tree;
$V$ is a finite set of non-terminal symbols, which is used to expand the scope of generation, representing intermediate states \ie, potential generation targets;
$\Sigma$ is a finite set of terminal symbols disjoint from $V$, indicating the actual contents of generation.
$R$ is a set of predefined rules that are used to iteratively transform the original SQLi payload into another equivalent form of SQLi payload that preserves the original functionality and maliciousness.

Taking the subtree of tautology (\ie, ``1 = 1'') in the hierarchical tree representation as an example, we can use $G$ to generate a set of candidate equivalent forms.
Alternatively, the tautology can be transformed into a boolean expression that turns out to be true $S_{True}$ (\ie, one of the non-terminal symbols) or terminal symbols, \eg, string $\tau$, complex $\tau$ and numeric $\tau$.
For the non-terminal $S_{True}$, we can either directly transform it into one of terminal symbols $\Sigma_{true}$ (\eg,  ``2$<>$3'',``True'', ``Not False''), or another non-terminal symbol in a recursive way, which could finally generate one of candidate SQLi payload ``true\textbackslash{}n\&\&/**foo*/select 1\textbackslash{}tand 2$<>$3''.

With the above tree representation and the CFG, in addition to covering the mutation operators in the baseline methods, we propose several novel and practical mutation operators (\eg, \textit{Inline Comment}, \textit{Where Rewriting}, \textit{DML Substitution}, and \textit{Tautology Substitution} in Table~\ref{table:examples-of-mutation-methods}).
With the following two examples, we stress that our semantic-based method is more general (\ie, covering more surface of payloads) and safe (\ie, semantic-preserving).
1) The existing methods can only identify ``1=1'' in payloads for mutation, while with the benefit of semantic-based matching, \mysystem can mutate for ``1=1'', ``1 = 1'', ``-3.7 = -3.7'', ``1 = 1.0'', ```foo'=`foo'''.
2) The existing methods mutate ``rlike'' to ``r='' and mutate ``order'' to ``$||$der'', to name a few, which will invalidate the entire payload.

\begin{algorithm}[htb!]
\small
\caption{Main Procedure of AdvSQLi}
\label{algo:AdvSQLi-main-procedure}
\SetKwFunction{BuildTree}{BuildTree}
\SetKwFunction{ExploreOperationalNodes}{ExploreOperationalNodes}
\SetKwFunction{CFG}{CFG}
\SetKwFunction{Min}{Min}
\SetKwFunction{MCTSState}{MCTSState}
\SetKwFunction{MCTSNode}{MCTSNode}
\SetKwFunction{MCTS}{MCTS}
\SetKwFunction{TreePolicy}{TreePolicy}
\SetKwFunction{DefaultPolicy}{DefaultPolicy}
\SetKwFunction{Backup}{Backup}
\SetKwFunction{BestChild}{BestChild}
\SetKwFunction{ConstructPayload}{ConstructPayload}
\SetKwInOut{Input}{Input}%
\SetKwInOut{Output}{Output}%
\Input{payload $x$, max steps $s$, computational $budget$, black-box situation $p$, WAF $clsf$.}
\Output{attack result, final score, final payload $x^{\prime}$.}
$t\leftarrow \BuildTree(x)$ \tcp*{Section \ref{sec:4.1}}
$t^{*} \leftarrow \ExploreOperationalNodes(t)$\;
$M \leftarrow \CFG(t^{*})$ \tcp*{Section \ref{sec:cfg}}
$s \leftarrow \Min(s, len(t^{*}))$ \tcp*{Max Steps.}
$state \leftarrow \MCTSState(t, M, p)$ \tcp*{Init State.}
$node \leftarrow \MCTSNode(state)$ \tcp*{Root Node.}
\For{$i\leftarrow 1$ \KwTo $s$}{
\For{$j\leftarrow 1$ \KwTo $budget$}{
$expand\_node \leftarrow \TreePolicy(node) $ \;
$reward \leftarrow \DefaultPolicy(expand\_node) $ \;
$\Backup(expand\_node, reward) $\;
}
$node^{\prime} \leftarrow \BestChild(node) $\;
$x^{\prime} \leftarrow \ConstructPayload(node^{\prime})$\;
$score \leftarrow clsf(x^{\prime})$\;
\If{$score < threshold\_of\_clsf $}{Return $True, score, x^{\prime}$\;}
$node \leftarrow node^{\prime}$\;
}
Return $False$;
\end{algorithm}

\subsection{MCTS Guided Search}\label{sec:search}
As our CFG-based mutation method can theoretically generate infinite candidates, we employ the Monte-Carlo tree search (MCTS) to solve it, which has proven effective in AlphaGo~\cite{silver2016mastering_alphago}.
In \mysystem, MCTS is to continuously build a search tree, where each node represents a state of the SQLi hierarchical tree, and the edges correspond to mutations, \ie, replacements of the node in the SQLi hierarchical tree. 
Commonly, MCTS contains four steps:
\textbf{Selection} is to find the best node worth exploring in the search tree employing the Upper Confidence Bounds (UCB) algorithm (Equation~\ref{equ:mcts-ucb}), \ie, the most worth exploring state of the SQL hierarchy tree.
\textbf{Expansion} is to perform an operation randomly and create a new child node, which means choosing a node in the hierarchical tree and replacing it with one of the equivalent replacements generated by CFG.
\textbf{Simulation} is to continue the ``game'' until reaching the maximum number of mutation steps.
Then, we can get its score.
\textbf{Back-propagation} is to provide feedback on the score of the newly expanded node to all the previous parent nodes, and update the score and visit times of these nodes to facilitate the calculation of the UCB score later.
\begin{equation}%
\footnotesize%
\label{equ:mcts-ucb}%
	score_{ucb} = \mathop{\arg\max}_{v^{\prime}\ \in\ children\ of\ v} (\frac{Q(v^{\prime})}{N(v^{\prime})}\ +\ c\sqrt{\frac{2\ln N(v)}{N(v^{\prime})}})
\end{equation}%

Besides, $v^{\prime}$ and $v$ represent the current node and its parent node respectively, $Q$ represents the cumulative quality value, $N$ represents the visit times, and $c$ is a parameter that can trade off between exploitation and exploration~\cite{kocsis2006bandit_mcts_c}.

The overall process of \mysystem is shown in Algorithm~\ref{algo:AdvSQLi-main-procedure}.
First, we represent the SQLi payload as a hierarchical tree $t$ (line 1), \ie, the first part of Figure~\ref{figure:pipeline-detail}.
After that, we find all operable nodes in $t$ (line 2), in other words, nodes such as database table names are set to a locked state.
We use CFG to generate equivalent nodes (or subtrees) for all operable nodes (line 3), \ie, the second part of Figure~\ref{figure:pipeline-detail}.
Then, we create the root node of MCTS, which contains the computational budget and our hierarchical tree (lines 5-6).
Further, we call MCTS within the max steps $s$ (lines 8-13), which is reflected in the third part of Figure~\ref{figure:pipeline-detail}.
According to the state after each round of the game, we can judge whether we win, \ie, reconstructing the SQLi payload based on the chain of all MCTS selections (line 14).
At last, we can obtain the bypassing payload.

\section{Evaluation}
\label{sec:5}
Next, we conduct an empirical evaluation of \mysystem.
We consider black-box settings with or without probability situations, which respectively correspond to state-of-the-art ML-based SQLi detectors and commercial WAF-as-a-service products. Under both settings, we compare \mysystem with a set of state-of-the-art attacks.
Note that due to their different optimization strategies (\eg, MCTS, RL, priority queue), it is challenging to find a unified configuration.
Instead, we adopt a two-stage approach that first assesses the validity of payloads and then conducts other evaluations.
Specifically, the evaluation is designed to answer the following key questions:

\noindent\textbf{$\bullet$ RQ1: Semantic preservation.}
\textit{Is} \mysystem \textit{able to preserve the original semantics of payloads?}

\noindent\textbf{$\bullet$ RQ2: Attack effectiveness against ML-based SQLi detection.}
\textit{Is} \mysystem \textit{effective against state-of-the-art ML-based SQLi detectors?}

\noindent\textbf{$\bullet$ RQ3: Attack effectiveness against WAF-as-a-service.}
\textit{Is} \mysystem \textit{effective against real-world WAF-as-a-service?}

\noindent\textbf{$\bullet$ RQ4: Ablation study.}
\textit{What is the effectiveness of each individual mutation method?}

\begin{figure}[htb!]
\centering
\subfigure[For \textbf{RQ1}: Employ the back-ends of various runtime environments to verify the functionality and maliciousness of generated payloads.]{
\label{subfigure:arch-1}
\includegraphics[width=0.49\textwidth]{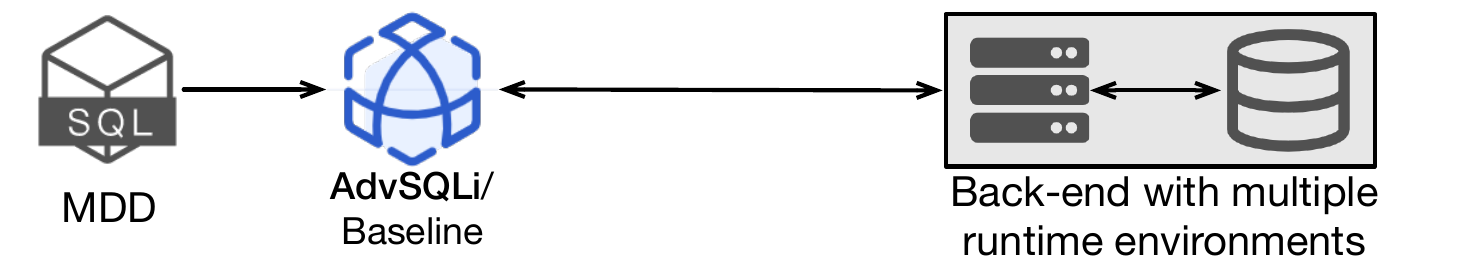}
}
\subfigure[For \textbf{RQ2}: Use ML-based SQLi detectors to evaluate the effectiveness and efficiency. The datasets are used in the training, validation and testing phases.]{
\label{subfigure:arch-2}
\includegraphics[width=0.49\textwidth]{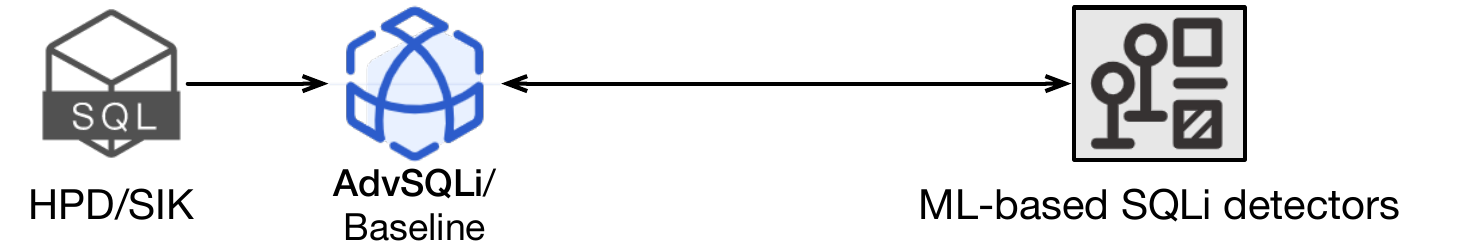}
}
\subfigure[For \textbf{RQ3} and \textbf{RQ4}: Evaluate the effectiveness when attacking commercial WAFs. The datasets are only used in the testing phase.]{
\label{subfigure:arch-3}
\includegraphics[width=0.49\textwidth]{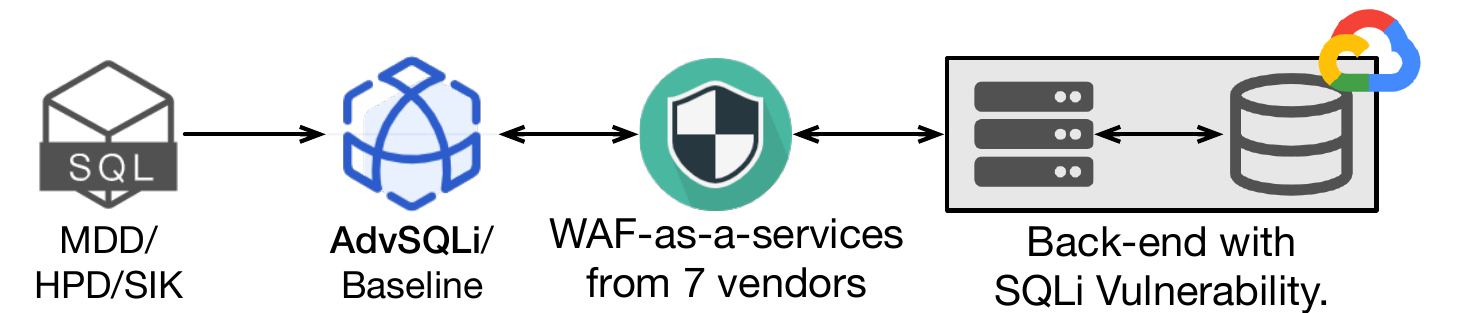}
}
\caption{Workflow of evaluation.}
\label{figure:arch}
\end{figure}

\noindent\textbf{Datasets.}
We evaluate the attack performance of \mysystem on two commonly used datasets \textbf{HPD} and \textbf{SIK}, and our own dataset \textbf{MDD}.
More precisely, \textbf{HPD}~\cite{website_hpd} is shorted for the HttpParamsDataset which not only includes all malicious SQLi payloads from CSIC~\cite{gimenez2010http_csic}, but also plenty of SQLi payloads generated by sqlmap~\cite{website_sqlmap}.
The two types of SQLi payloads in HPD are widely used in SQLi detection~\cite{ring2019survey_csic_ref1,parra2020detecting_csic_ref2,luo2020novel_csic_ref3}.
\textbf{SIK}~\cite{website_sik} is shorted for the SQL injection dataset in Kaggle, which is employed as the evaluation dataset in one of our baseline attack methods.
For each dataset, as the number of benign samples in each dataset is about twice that of malicious samples, we randomly select 3,000 samples (2,000 benign samples and 1,000 malicious samples) as the validation set and the testing set, respectively, and make the remaining samples as the training set.
However, due to the diversity of samples in the above datasets, we cannot build corresponding runtime environments in a short time.
In response, to verify the functionality and maliciousness of the generated payloads dynamically, we manually constructed \textbf{MDD}, including prevalent SQLi payloads with union-based injection, blind injection, error-based injection, and so on.

\begin{table}[htb!]
    \centering
    \small
    \caption{\upshape Summary statistics of datasets. The number before or after ``/'' denotes the number of benign/malicious payloads.}
    \begin{tabular}{ccccc}
    \toprule
    Datasets & Training & Validation & Testing & Total \\
    \midrule
    \textbf{HPD} & 15,304/8,852 & 2,000/1,000 & 2,000/1,000 & 30,156 \\ 
    \textbf{SIK} & 12,840/9,168 & 2,000/1,000 & 2,000/1,000 & 28,008 \\ 
    \textbf{MDD} & - & - & -/100 & 100 \\ 
    \bottomrule
    \end{tabular}
    \label{table:dataset_MMD}
\end{table}

\noindent\textbf{Target Models.}
The target WAFs are divided into two categories to cover black-box with and without probability situations:

\textbf{1) SQLi Detection Models.}
We use the character-based anomaly detection model WAF-Brain~\cite{website_wafbrain} as one of our target models, which is a state-of-the-art AI-based WAF and is widely used by previous work~\cite{demetrio2020waf_wafamole,wang2020evading_gymwaf}.
However, given the unsatisfactory generalization of WAF-Brain on our datasets, we train two classification models (\ie, CNN~\cite{zhang2015sensitivity_cnn2,mikolov2017advances_cnn3,rakhlin2016convolutional_cnn1} and LSTM~\cite{zhang2019machine_lstm1,yu2018attention_lstm2}) based on state-of-the-art SQLi detection and text classification literature.
We use ``='', ``$|$'', ``,'' and ``\textvisiblespace'' as keywords for vocabulary segmentation and train both CNN and LSTM models at the word level on HPD and SIK datasets using PyTorch.
The performance of the three models is shown in Table~\ref{table:models-perforce}.
As subsequent evaluations show that the WAF-Brain is too easy to bypass at FPR of 1\% and 1\perthousand, we manually designate a threshold of 0.1 following the settings of ~\cite{wang2020evading_gymwaf}.

\begin{table}[htb!]
\centering
\caption{\upshape Performance of target models. $\delta$ means the threshold calculated by FPR.}
\begin{tabular}{ccccccc}
\toprule
\multirow{2}[3]{*}{\begin{tabular}[c]{@{}c@{}}Target\\ Model\end{tabular}} & \multirow{2}[3]{*}{Dataset} & \multirow{2}[3]{*}{\begin{tabular}[c]{@{}c@{}}AUC\\ (\%)\end{tabular}} & \multicolumn{2}{c}{FPR=1\%} & \multicolumn{2}{c}{FPR=1\perthousand} \\
\cmidrule(r{1pt}){4-5} \cmidrule(l{1pt}){6-7}
 &  &  & Acc (\%) & $\delta$ & Acc (\%) & $\delta$ \\
\midrule
$\text{WAF-Brain}$ & HPD & 99.14 & 85.80 & 0.333 & 62.18 & 0.500 \\
$\text{WAF-Brain}$ & SIK & 96.97 & 68.45 & 0.286 & 54.25 & 0.572 \\
$\text{CNN}$ & HPD & 99.96 & 99.45 & 0.044 & 99.90 & 0.159 \\
$\text{CNN}$ & SIK & 99.99 & 99.45 & 0.007 & 99.85 & 0.145 \\
$\text{LSTM}$ &HPD & 99.96 & 99.45 & 0.002 & 99.90 & 0.006 \\
$\text{LSTM}$ & SIK & 99.99 & 99.50 & 0.025 & 99.90 & 0.105 \\
 \bottomrule
\end{tabular}
\label{table:models-perforce}
\end{table}

\textbf{2) Real-world WAF-as-a-service Solutions.}
To verify the real-world attack effectiveness of \mysystem, we purchase and deploy 7 WAF-as-a-service solutions from mainstream vendors including Amazon Web Services (AWS), F5, Cyber Security Cloud (CSC), Fortinet, Cloudflare, Wallarm, and the state-of-the-art open-sourced WAF ModSecurity.
Specifically, we employ the following real-world settings to configure and deploy them:
\textbf{AWS, F5, CSC, Fortinet:} We create four Access Control Lists (ACLs) on AWS.
Further, we subscribe to the rules of these vendors and integrate them into the corresponding ACL.
\textbf{Cloudflare:} We subscribe to its pro plan to enable the full-blown WAF.
\textbf{Wallarm:} We deploy a Wallarm node on the Google Cloud Platform.
\textbf{ModSecurity:} We build ModSecurity based on Nginx and embed the latest version of the OWASP CoreRuleSet (CRS) in it.

\noindent\textbf{Attack Methods.}
Previous research has made some progress in generating adversarial SQLi payloads.
1) \mole: Demetrio \etal~\cite{demetrio2020waf_wafamole} defined 7 mutation methods based on regular expression, and used a priority queue to guide their mutation process (\ie, payloads with low scores are used as the initial payloads for the next round).
2) Wang \etal~\cite{wang2020evading_gymwaf} and Hemmati \etal~\cite{hemmati2021using} followed the mutation operators in~\cite{demetrio2020waf_wafamole}, and they proposed a search strategy based on RL~\cite{sutton2018reinforcement_suttonrl}.
We use \mole as one of the baseline methods.
Since neither~\cite{wang2020evading_gymwaf} nor~\cite{hemmati2021using} has open-sourced attack modules, we implement their method based on DQN and call it \rl. 
In addition, to evaluate the effectiveness of the MCTS method and to explore the upper limit of attack, we implement two variants of \mysystem:
1) \mysystemr, which randomly combines the candidate nodes of the hierarchical tree.
2) \mysystemv, which performs permutations on the candidates of all nodes.

\begin{table}[htb!]
\centering
\caption{\upshape Attack methods. ``Both'' means it is suitable for both black-box with and without probability situations, \ie, ML-based SQLi detection and real-world WAF-as-a-service.}
\begin{tabular}{llll}
\hline
Method & Capability & Mutate Method & Search Strategy \\
\hline
\mole & BB w/ prob. & String-based & Priority Queue \\
\rl & BB w/ prob. & String-based & RL (DQN) \\
\textbf{\textbf{\mysystemm}} & \textbf{Both} & \textbf{Semantic-based} & \textbf{MCTS} \\
\mysystemr & Both & Semantic-based & Random \\
\mysystemv & Both & Semantic-based & Violent \\
\hline
\end{tabular}
\label{table:attack-methods}
\end{table}

\noindent\textbf{Metrics and Parameters.}
The metrics of our attack evaluations are:
(\romannumeral1) \textbf{Attack Success Rate (ASR)} represents the percentage of malicious samples that can bypass the WAF after the attack.
(\romannumeral2) \textbf{Validity of Generated Payloads (VGP)} reflects the feature of semantic-preserving of an attack method.
(\romannumeral3) \textbf{Query} refers to the query number we perform on a payload to bypass WAF.
Here are some parameters that may affect the running results:
In \mysystem and \mole, the key parameters are \textbf{Step} and \textbf{Budget}.
Recall from Algorithm~\ref{algo:AdvSQLi-main-procedure} that the outer loop (lines 6-16) is constrained by Step, which repetitively builds a search tree and updates the optimal \textit{node}.
Moreover, the inner loop (lines 7-10) is constrained by Budget, which iteratively builds and extends the search tree under the current root \textit{node}.
As for \mole, we treat the size of each priority queue as the budget and the number of priority queue rounds as a step.
Therefore multiple queries are performed in a step in \mole and \mysystem.
Besides, the number of steps and queries are equal in \rl and \mysystemr.

\subsection{RQ1: Semantic Preservation }
\noindent\textbf{Evaluation setup.}
We propose a dynamic method to perform the semantic-preserving assessment, \ie, comparing the running results of the generated payloads and the original one.
Figure~\ref{subfigure:arch-1} sketches the workflow.
Specifically, we first create two databases (MySQL 5.7 and MySQL 8.0) and build the corresponding tables, columns, and data based on MDD.
Then, we integrate these databases in back-end scripts (PHP 5.6, PHP 7.4, and Python 3.7) with SQLi vulnerabilities.
This leaves us with 6 web services where we can steal information through SQLi attacks.
Furthermore, we mutate the samples in MDD utilizing \mysystem and \mole, and then send the mutated payloads to these web services to observe the running results.
Since the mutation modules of \mole and \rl are identical, we only evaluate \mole.
In this context, we evaluate the relationship between VGP and attack steps.
Roughly, we generate about 10,000 mutated payloads with \mole and \mysystem, respectively.
Moreover, to determine the subsequent parameters for a fair comparison, we evaluate the relationship between ASR and attack steps of \mole based on multiple models.

\begin{figure}[htb!]
\centering
\includegraphics[width=0.45\textwidth,keepaspectratio=true]{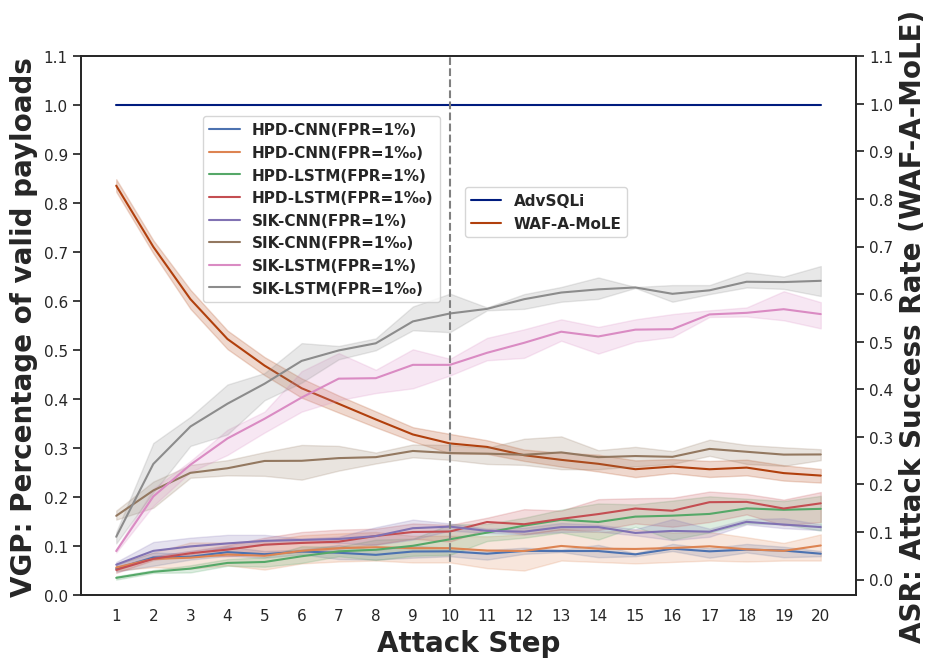}
\caption{VGP of \mysystem and relationship between VGP and ASR of \mole. The horizontal blue line means that the VGP of \mysystem is always 100\%. The descending orange line refers to the changing trend of the VGP of \mole with the number of attack steps, which is the average value under various runtime environments. The rising lines refer to the changing trend of the ASR with the number of attack steps.}
\label{figure:valid}
\end{figure}

\noindent\textbf{Results.}
From Figure~\ref{figure:valid}, we can see that the payloads mutated by \mysystem are always valid (VGP is 1.0), \ie, preserving the original functionality and maliciousness.
This is inseparable from our semantic-based mutation method.
However, the string-based mutation method in \mole and \rl mutates ``rlike'' to ``r='' and mutate ``order'' to ``$||$der'', to name a few, which do damage the original semantics of the payloads and invalidate them.
For example, half of the mutated payloads generated by the string-based mutation method are invalid when the number of attack steps is 4.
Besides, Figure~\ref{figure:valid} shows that both the VGP and ASR curves of \mole tend to be flat when the attack step is larger than 10.
For a fair comparison, we take the number of attack steps at 10 as the default parameter of \mole for the subsequent evaluations. 

\begin{center}
\fcolorbox{black}{gray!5}{\parbox{0.95\linewidth}{
\textbf{Answer to RQ1: }\mysystem is semantic-preserving, while the baseline methods (\ie, \mole and \rl) fail to preserve semantics. Namely, the adversarial payloads generated by \mysystem maintain the original functionality and maliciousness.}}
\end{center}

\begin{table*}[htb]
\centering
\setlength\tabcolsep{1pt}
\small
\caption{\upshape The ASR results of attacking against \local when under default settings and when increasing the number of attack steps to 20. \texttt{WAM} means \mole. $\delta$ means the threshold calculated by FPR.}
\begin{tabular}{cccccccccccccc}
\toprule
\multirow{3}[4]{*}{Dataset} & \multirow{3}[6]{*}{\tabincell{c}{Target SQLi\\Detection}} & \multirow{3}[4]{*}{$\delta$} & \multirow{3}[4]{*}{\tabincell{c}{FNR\\(\%)}} &   \multicolumn{9}{c}{Attack Success Rate (\%)}\\
\cmidrule{5-14}          &       &       &       & \multicolumn{5}{c}{Parameter: Step = 10 (Default)} & \multicolumn{5}{c}{Parameter: Step = 20} \\
\cmidrule(r{1pt}){5-9} \cmidrule(l{1pt}){10-14}          &       &       &       & \texttt{WAM}      & \rl   & \mysystemr & \mysystem     & \mysystemv     & \texttt{WAM}     & \rl    & \mysystemr    & \mysystemm    & \mysystemv \\
\midrule
\multirow{7}{*}{HPD} & $\text{WAF-Brain}^{\delta=0.1}$ & 0.100 & 0.1 & 75.56 & 4.15 & 97.34 & \textbf{99.97} & 98.01 & 99.86 & 8.62 & 97.47 & \textbf{99.93} & 98.04 \\
 & $\text{WAF-Brain}^{1\%}$ & 0.333 & 31.3 & 100 & 69.62 & 100 & \textbf{100} & 99.59 & 100 & 86.15 & 100 & \textbf{100} & 100 \\
 & $\text{WAF-Brain}^{1\perthousand}$ & 0.500 & 76.7 & 100 & 81.95 & 100 & \textbf{100} & 100 & 100 & 90.40 & 100 & \textbf{100} & 100 \\
 & $\text{CNN}^{1\%}$ & 0.044 & 0.1 & 9.34 & 6.97 & 7.94 & \textbf{26.46} & 25 & 11.48 & 7.54 & 9.58 & \textbf{27.03} & 26.12 \\
 & $\text{CNN}^{1\perthousand}$ & 0.159 & 0.1 & 13.68 & 9.44 & 13.28 & \textbf{34.90} & 32.9 & 16.88 & 10.54 & 14.78 & \textbf{36.27} & 34.93 \\
 & $\text{LSTM}^{1\%}$ & 0.002 & 0.1 & 10.18 & 3.90 & 9.34 & \textbf{23.62} & 16.8 & 14.41 & 4.40 & 11.34 & \textbf{23.99} & 20.32 \\
 & $\text{LSTM}^{1\perthousand}$ & 0.006 & 0.1 & 11.41 & 5.04 & 11.08 & \textbf{26.03} & 18.2 & 15.85 & 5.47 & 13.08 & \textbf{26.43} & 22.72 \\
 \midrule
\multirow{7}{*}{SIK} & $\text{WAF-Brain}^{\delta=0.1}$ & 0.100 & 0.5 & 77.12 & 13.89 & 98.06 & \textbf{100} & 99.16 & 99.93 & 24.40 & 99.48 & \textbf{100} & 100 \\
 & $\text{WAF-Brain}^{1\%}$ & 0.286 & 62.5 & 99.93 & 65.31 & 99.93 & \textbf{100} & 99.64 & 100 & 82.15 & 100 & \textbf{100} & 100 \\
 & $\text{WAF-Brain}^{1\perthousand}$ & 0.572 & 91.5 & 100 & 98.85 & 100 & \textbf{100} & 100 & 100 & 100 & 100 & \textbf{100} & 100 \\
 & $\text{CNN}^{1\%}$ & 0.007 & 0.1 & 25.16 & 11.01 & 12.48 & \textbf{35.97} & 27 & 29 & 12.98 & 15.55 & \textbf{39.74} & 32.23 \\
 & $\text{CNN}^{1\perthousand}$ & 0.145 & 0.2 & 44.12 & 25.05 & 27.42 & \textbf{56.78} & 45 & 49.30 & 27.25 & 35.64 & \textbf{58.48} & 52.40 \\
 & $\text{LSTM}^{1\%}$ & 0.025 & 0.0 & 36.50 & 9.87 & 20.57 & \textbf{78.80} & 61.40 & 49.53 & 10.77 & 31.03 & \textbf{83.10} & 74.20 \\
 & $\text{LSTM}^{1\perthousand}$ & 0.105 & 0.1 & 42.61 & 12.45 & 27.53 & \textbf{91.36} & 74.90 & 56.52 & 13.58 & 39.87 & \textbf{95.03} & 84.68 \\
\bottomrule
\end{tabular}
\label{table:attack-against-models-results}
\end{table*}

\subsection{RQ2: Attack Effectiveness against ML-based SQLi Detection}
\noindent\textbf{Evaluation setup.}
We evaluate the attack effectiveness and efficiency by utilizing the attack methods in Table~\ref{table:attack-methods} to attack the models in Table~\ref{table:models-perforce}.
We set the maximum number of steps of each attack method to 10 and set the default budget of \mysystem and \mole to 10.
Besides, we set the query number of attempts of \mysystemr to 10,000 for efficiency.

\begin{figure}[htb!]
\centering
\subfigure[HPD-LSTM(FPR=1\%)]{
\label{subfigure:asr-query-2}
\includegraphics[width=0.22\textwidth]{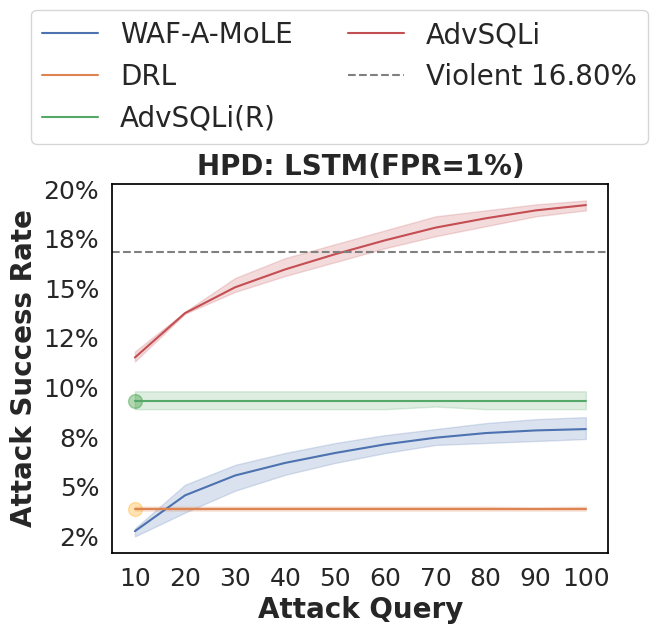}
}
\subfigure[SIK-CNN(FPR=1\%)]{
\label{subfigure:asr-query-3}
\includegraphics[width=0.22\textwidth]{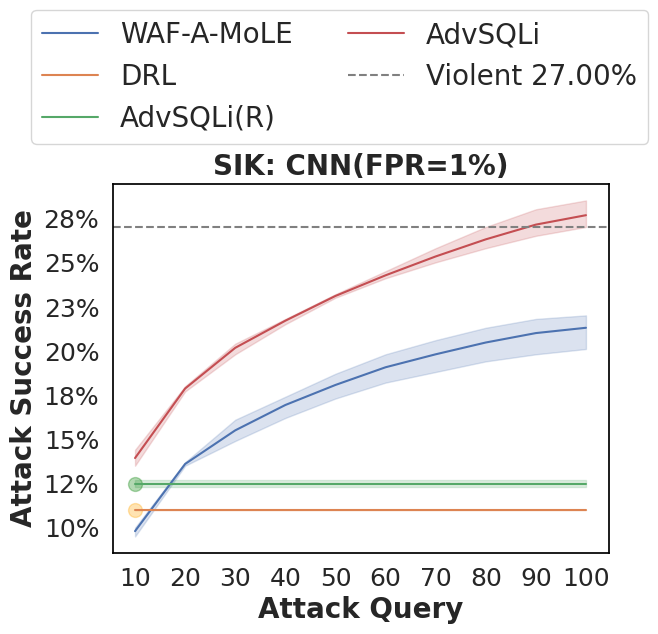}
}
\caption{Relationships between ASR and Query (within 100) when attacking against \local. The grey dotted line results from the \mysystemr under the default settings.}
\label{figure:asr-query}
\end{figure}

\noindent\textbf{Results.}
\textbf{1) Effective.}
The middle part of Table~\ref{table:attack-against-models-results} shows the ASR results of attacking against \local under default settings.
Intuitively, \local based on anomaly detection, CNN, and LSTM are all vulnerable to adversarial attacks.
For example, several attack methods can achieve ASRs of 100\% on WAF-Brain.
We find that \mysystem is better than other attack methods in terms of higher ASRs.
For instance, the ASR of \mysystem is twice that of \mole, three times that of \mysystemr, and seven times that of \rl when attacking against the LSTM model (FPR=1\perthousand) under the SIK dataset.
\textbf{2) Efficient.} Figure~\ref{figure:asr-query} sketches the relationships between ASR and Query.
It is clear that \mysystem always outperforms other methods, regardless of the number of queries.
In most cases, the ASRs of \mysystem surpass \mysystemr within dozens of queries.
Compared with other baseline methods, the effectiveness of \mole gradually surpasses others with the increase of queries.
However, it still tends to be stable in the end and cannot exceed \mysystem.

\noindent\textbf{Parameters adjustment.}
We conduct detailed evaluations to further explore the impact of the parameters mentioned above on the effectiveness and efficiency of different attack methods.
We first increase the maximum number of steps of \mysystem and \mole to 20 and repeat all evaluations.
In order to explore the impact of the computational budget on the attack results, we increase the budget of \mysystem and \mole to 20.

\noindent\textbf{Results.}
\textbf{1) Step.}
The results of ASR are shown in the right part of Table~\ref{table:attack-against-models-results}.
Evidently, the ASRs of various attack methods increase more or less.
However, \mysystem still has an absolute advantage over the baseline methods. 
\textbf{2) Budget.}
From Table~\ref{table:attack-against-models-results-budget}, we can observe that the ASRs improve as the budget increases for almost all models.
Moreover, when the budget is 20, \mole is still not better than \mysystem with budget 10, which proves the advantage of \mysystem.
\begin{center}
\fcolorbox{black}{gray!5}{\parbox{0.95\linewidth}{
\textbf{Answer to RQ2: }\mysystem is more effective and efficient than all baseline methods, achieving higher ASRs with fewer queries.}}
\end{center}

\begin{table}[htb!]
\small
\setlength\tabcolsep{3pt}
\centering
\caption{\upshape The ASR results under budget is 10 and 20. WAM means WAF-A-MoLE}
\begin{tabular}{cccccc}
\toprule
\multirow{3}[4]{*}{Dataset} & \multirow{3}[4]{*}{\begin{tabular}[c]{@{}c@{}}Target SQLi \\ Detection\end{tabular}} & \multicolumn{4}{c}{Attack Success Rate (\%)} \\
\cmidrule{3-6}
 &  & \multicolumn{2}{c}{Budget=10} & \multicolumn{2}{c}{Budget=20} \\
 \cmidrule(r{1pt}){3-4} \cmidrule(l{1pt}){5-6}
 &  & \texttt{WAM} & \mysystem & \texttt{WAM} & \mysystem \\
 \midrule
\multirow{7}{*}{HPD} & $\text{WB}^{\delta=0.1}$ & 75.56 & 99.97 & 99.56 & \textbf{100} \\
 & $\text{WB}^{1\%}$ & 100 & 100 & 100 & 100 \\
 & $\text{WB}^{1\perthousand}$ & 100 & 100 & 100 & 100 \\
 & $\text{CNN}^{1\%}$ & 9.34 & 26.46 & 10.54 & \textbf{27.83} \\
 & $\text{CNN}^{1\perthousand}$ & 13.68 & 34.90 & 14.71 & \textbf{36.87} \\
 & $\text{LSTM}^{1\%}$ & 10.18 & 23.62 & 11.58 & \textbf{24.59} \\
 & $\text{LSTM}^{1\perthousand}$ & 11.41 & 26.03 & 12.21 & \textbf{27.09} \\
 \midrule
\multirow{7}{*}{SIK} & $\text{WB}^{\delta=0.1}$ & 77.12 & 100 & 99.59 & \textbf{100} \\
 & $\text{WB}^{1\%}$ & 99.93 & 100 & 100 & 100\\
 & $\text{WB}^{1\perthousand}$ & 100 & 100 & 100 & 100 \\
 & $\text{CNN}^{1\%}$ & 25.16 & 35.97 & 26.49 & \textbf{38.64}\\
 & $\text{CNN}^{1\perthousand}$ & 44.12 & 56.78 & 46.62 & \textbf{58.22}\\
 & $\text{LSTM}^{1\%}$ & 36.50 & 78.8 & 41.23 & \textbf{82} \\
 & $\text{LSTM}^{1\perthousand}$ & 42.61 & 91.36 & 48.21 & \textbf{94.49} \\
 \bottomrule
\end{tabular}
\label{table:attack-against-models-results-budget}
\end{table}
\vspace{-2mm}

\subsection{RQ3: Attack Effectiveness against WAF-as-a-service}
\noindent\textbf{Evaluation setup.}
We deploy a modified version of SQLi-labs~\cite{website_sqlilabs} with the underlying database on the Google Cloud Platform.
Further, we protect it by utilizing seven real-world WAFs in turn by modifying DNS resolutions and forwarding traffic.
In this case, \mysystem acts as a client to continuously send attack requests to the protected web service.

\noindent\textbf{Real-world adaptation.}
Note that, our modified back-end web service supports 4 common HTTP request methods (\ie, GET, GET(JSON), POST, and POST(JSON)) corresponding to comprehensive real-world situations.
We fine-tune the mutation process in \mysystem adaptively to preserve the semantics of the generated payloads across different request methods, \eg, \textbf{\textbackslash{}n} needs to be encoded as \textbf{\%0A} for non-JSON requests and the ``\textbf{\#}'' sign cannot be present in comments (\textbf{/*\#*/}) when sending the payload via GET.
As for the baseline methods, the mutation methods of \mole and \rl are identical, and do not support attacking black-box WAFs given their score-based attack guidance.
Therefore, we implement a random search-based variant for \mole and perform the same adaptations in mutation methods.
We employ \mole to attack the web service protected by ModSecurity and elide the results of other WAF-as-a-service for experimental efficiency.

\begin{table}[htb!]
\centering
\setlength\tabcolsep{1pt}
\small
\caption{\upshape Results of attacking against real-world WAFs. A(R) means AdvSQLi(R).}
\begin{tabular}{cccccccc}
\toprule
\multirow{3}[4]{*}{WAFaaS} & \multirow{3}[4]{*}{\begin{tabular}[c]{@{}c@{}}Request\\ Method\end{tabular}} & \multicolumn{3}{c}{HPD} & \multicolumn{3}{c}{SIK} \\
\cmidrule(l{1pt}r{1pt}){3-5} \cmidrule(l{1pt}r{1pt}){6-8}
 &  & \multirow{2}{*}{\begin{tabular}[c]{@{}c@{}}FNR\\ (\%)\end{tabular}} & \multicolumn{2}{c}{ASR(\%)} & \multirow{2}{*}{\begin{tabular}[c]{@{}c@{}}FNR\\ (\%)\end{tabular}} & \multicolumn{2}{c}{ASR(\%)} \\
 \cmidrule(r{1pt}){4-5} \cmidrule(l{1pt}){7-8}
 &  &  & \texttt{A(R)} & \texttt{AdvSQLi} &  & \texttt{A(R)} & \texttt{AdvSQLi} \\
 \midrule
\multirow{4}{*}{AWS} & GET & 5.3 & 15.21 & \textbf{18.69} & 8.2 & 10.80 & \textbf{14.39} \\
 & GET(JSON) & 60.2 & 86.43 & \textbf{89.45} & 63.4 & 96.45 & \textbf{99.73} \\
 & POST & 3.4 & 29.19 & \textbf{30.02} & 14.5 & 26.46 & \textbf{31.97} \\
 & POST(JSON) & 60.2 & 84.17 & \textbf{89.45} & 63.4 & 96.17 & \textbf{99.73} \\
 \midrule
\multirow{4}{*}{F5} & GET & 40.7 & 70.66 & \textbf{82.46} & 45.1 & 69.95 & \textbf{79.60} \\
 & GET(JSON) & 40.5 & 67.73 & \textbf{83.87} & 43.7 & 61.99 & \textbf{82.06} \\
 & POST & 35.6 & 70.50 & \textbf{83.7} & 41.9 & 66.09 & \textbf{80.72} \\
 & POST(JSON) & 35.4 & 71.05 & \textbf{85.76} & 40.5 & 60.50 & \textbf{82.69} \\
 \midrule
\multirow{4}{*}{\begin{tabular}[c]{@{}c@{}}Cyber\\Security\\Cloud\end{tabular}} & GET & 19.7 & 63.14 & \textbf{77.33} & 37.1 & 50.08 & \textbf{70.27} \\
 & GET(JSON) & 20 & 65.75 & \textbf{77.38} & 37.1 & 53.26 & \textbf{70.91} \\
 & POST & 19.7 & 63.26 & \textbf{75.22} & 37.1 & 46.97 & \textbf{70.38} \\
 & POST(JSON) & 20 & 64.50 & \textbf{74.50} & 37.1 & 52.46 & \textbf{71.38} \\
 \midrule
\multirow{4}{*}{Fortinet} & GET & 8.8 & 48.25 & \textbf{53.40} & 14.2 & 45.16 & \textbf{55.19} \\
 & GET(JSON) & 9.7 & 78.07 & \textbf{83.17} & 15.7 & 73.40 & \textbf{81.24} \\
 & POST & 8.8 & 48.9 & \textbf{53.40} & 14.0 & 45.75 & \textbf{55.06} \\
 & POST(JSON) & 9.7 & 77.52 & \textbf{83.17} & 15.5 & 73.58 & \textbf{81.28} \\
 \midrule
\multirow{4}{*}{\begin{tabular}[c]{@{}c@{}}Cloud-\\ flare\end{tabular}} & GET & 8.1 & 20.13 & \textbf{21.33} & 18.8 & 26.39 & \textbf{32.43} \\
 & GET(JSON) & 17.7 & 35.60 & \textbf{37.79} & 29.2 & 55.73 & \textbf{58.13} \\
 & POST & 47.1 & 35.35 & \textbf{35.92} & 63.2 & 47.96 & \textbf{48.77} \\
 & POST(JSON) & 47.1 & 35.16 & \textbf{35.92} & 63.2 & 47.96 & \textbf{49.05} \\
 \midrule
\multirow{4}{*}{Wallarm} & GET & 1.4 & 16.94 & \textbf{18.76} & 6.5 & 24.20 & \textbf{33.94} \\
 & GET(JSON) & 1.4 & 16.94 & \textbf{18.76} & 6.4 & 24.17 & \textbf{34.01} \\
 & POST & 1.4 & 15.14 & \textbf{17.28} & 6.7 & 20.52 & \textbf{31.26} \\
 & POST(JSON) & 2.4 & 16.79 & \textbf{17.40} & 7.6 & 23.86 & \textbf{32.24} \\
 \midrule
\multirow{4}{*}{\begin{tabular}[c]{@{}c@{}}Mod-\\ Security\end{tabular}} & GET & 0.1 & 5.74 & \textbf{11.61} & 3.3 & 5.63 & \textbf{10.88} \\
 & GET(JSON) & 20.1 & 42.59 & \textbf{49.06} & 30.9 & 49.73 & \textbf{58.32} \\
 & POST & 0.1 & 5.44 & \textbf{10.61} & 3.5 & 4.53 & \textbf{9.55} \\
 & POST(JSON) & 0.1 & 4.90 & \textbf{10.61} & 3.5 & 4.50 & \textbf{9.55} \\
 \bottomrule
\end{tabular}
\label{table:attack-against-realwaf}
\end{table}

\noindent\textbf{Results.}
Table~\ref{table:attack-against-realwaf} illustrates the FNR and ASR results of attacking against 7 WAF-as-a-service, and Table~\ref{table:asr-and-query-against-mods} shows the ASR and Query results of \mysystem and \mole against ModSecurity.

\textbf{1) FNR} directly reflects the protection effectiveness of WAFs.
In general, the protective effects of each WAF are uneven.
Classified according to different request methods, in the majority of cases, the FNRs of the two GET-based request methods are greater than or equal to those for POST-based methods, which means that the protection of the latter is stricter.
If this is not a design flaw, it is likely that the vendor deliberately did it.
When making requests to web services via POST, there is typically a larger attack payload, leading vendors to implement stricter protective measures.
In contrast, GET requests, which generally do not carry such complex content, may be subject to more lenient scrutiny to prevent the accidental interception of legitimate traffic.
Further, we can divide WAFs into four categories by inferring WAF's responses to different requests based on FNRs:
\textbf{1) }F5, CSC, Fortinet, and Wallarm treat the four request methods equally.
\textbf{2) }Cloudflare implements different strategies based on whether the request method is GET or POST.
\textbf{3) }AWS treats the payload separately based on whether the request parameter is in JSON type.
\textbf{4) }ModSecurity processes requests via GET(JSON) separately, yet treats the remaining three request methods equally.
Alarmingly, both AWS and Cloudflare have FNRs of over 60\% on the SIK dataset.
In other words, more than 60\% of malicious requests will be directly forwarded to the back-end service.
On the contrary, Fortinet and Wallarm perform well as their FNRs are low and relatively even among different request methods.

\textbf{2) ASR} illustrates the robustness of a WAF against adversarial attacks.
If a WAF exhibits high FNRs and ASRs simultaneously, it is evident that its protective effect is unsatisfactory.
Overall, our attacks against WAF-as-a-service are successful.
The ASRs against AWS on the SIK dataset even reached 99.73\%.
Besides, the ASRs of each WAF have a similar distribution to FNRs, which further confirms our conclusion that we have drawn above on the effectiveness of WAFs.
By analyzing the ASRs, we can find that when requesting web services in JSON-type parameters, ASRs are much higher than non-JSON.
It is a great security risk given the current development trend of web services, \ie, more and more web services are designed based on the decoupled architecture, most of which use JSON to transfer data.

\begin{table}[htb!]
\small
\setlength\tabcolsep{3pt}
\centering
\caption{\upshape The ASR and Query results of \mysystem and \mole against ModSecurity. The average query is calculated based on successful samples crossed in both attack methods. WAM means WAF-A-MoLE.}
\begin{tabular}{cccccc}
\toprule
\multirow{2}{*}{Dataset} & \multirow{2}{*}{Request Method} & \multicolumn{2}{c}{ASR(\%)} & \multicolumn{2}{c}{Average Query} \\
\cmidrule(r{1pt}){3-4} \cmidrule(l{1pt}){5-6}
 &  & AdvSQLi & WAM & AdvSQLi & WAM\\
  \midrule
\multirow{4}{*}{HPD} & GET & \textbf{11.61} & 6.61 & \textbf{9.80} & 30.11 \\
 & GET(JSON) & \textbf{49.06} & 47.02 & \textbf{10.15} & 18.91 \\
 & POST & \textbf{10.61} & 5.61 & \textbf{10.66} & 29.66 \\
 & POST(JSON) & \textbf{10.61} & 5.31 & \textbf{9.93} & 30.31 \\
    \midrule
\multirow{4}{*}{SIK} & GET & \textbf{10.88} & 7.34 & \textbf{10.67} & 31.5 \\
 & GET(JSON) & \textbf{58.32} & 48.05 & \textbf{10.26} & 32.65 \\
 & POST & \textbf{9.55} & 7.36 & \textbf{10.37} & 30.44 \\
 & POST(JSON) & \textbf{9.55} & 5.80 & \textbf{10.65} & 30.32 \\
 \bottomrule
\end{tabular}
\label{table:asr-and-query-against-mods}
\end{table}

The four WAFs hosted on AWS (AWS, F5, CSC, and Fortinet) are less capable of preventing SQLi than other WAFs.
Wallarm is effective because it has low FNRs and ASRs.
Besides, Fortinet has ASRs many times higher than FNRs, which means that it cannot defend against adversarial attacks very well.
Interestingly, ModSecurity shows the strongest robustness if we ignore the shortcomings in the defense of GET(JSON) requests, which means that open-sourced tools may have natural advantages.
Comparing the two attack methods we proposed, overall, \mysystemm has much higher ASRs than \mysystemr against all WAF-as-a-service. 
We can see from Table~\ref{table:attack-against-realwaf} that in hard-to-attack situations (such as ModSecurity), the ASRs of \mysystem are twice that of \mysystemr.
From Table~\ref{table:asr-and-query-against-mods}, we can see that \mysystem is more effective and efficient than \mole, as it achieves nearly twice the ASRs with only one-third queries of \mole in several cases.

\begin{center}
\fcolorbox{black}{gray!5}{\parbox{0.99\linewidth}{
\textbf{Answer to RQ3: }All WAF-as-a-service under tests are vulnerable to \mysystem. More importantly, some vendors have severe deficiencies, \eg, the insufficiency in parsing JSON-type parameters. \mysystem is more effective than the baseline methods when attacking against real-world WAF-as-a-service.}}
\end{center}

\subsection{RQ4: Ablation Study}
\label{ablation}
\begin{figure}[htb!]
    \centering
    \includegraphics[width=0.49\textwidth]{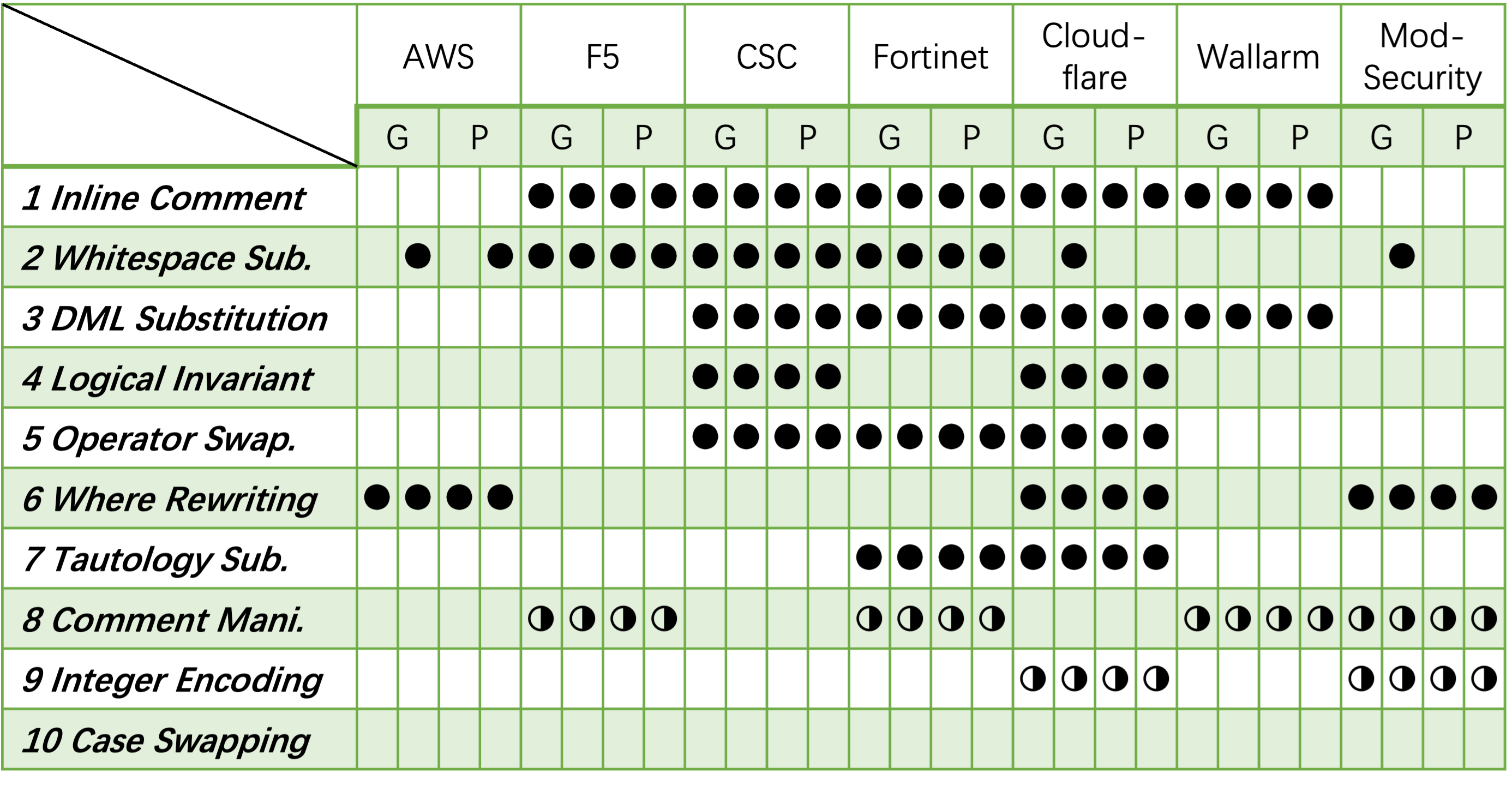}
    \caption{Effective mutation methods in bypassing WAF-as-a-service. Full circle means that this action is effective in most cases; Half circle indicates that it may work under specific payloads or combined with other mutations.}
    \label{table:valid-actions}
\end{figure}

\noindent\textbf{Overall conclusion.}
In addition to analyzing the ASR results, it is meaningful to know the root causes of bypassing, \ie, the concrete vulnerabilities of WAFs.
We automatically analyze all bypass samples in RQ3 based on the hierarchical trees to conclude more comprehensively.
Supplemented by manual methods, we get the conclusions in Figure~\ref{table:valid-actions} within the format of Table~\ref{table:examples-of-mutation-methods}.
Intuitively, individual attack methods are not always effective in attacking all WAF-as-a-service.
The mutation methods we propose (\ie, \textit{Inline Comment} and \textit{DML Substitution}) and \textit{Whitespace Substitution} are the most effective methods.
Interesting, just adding some comments into the SQLi payloads can bypass F5, Fortinet, Wallarm, and ModSecurity in some cases.
We speculate that the detection signatures in these WAFs are not robust, causing these bypasses.

\noindent\textbf{Case studies.}
Hereafter, we further analyze the vulnerabilities through specific examples.
We select three representative payloads from MDD, which can steal table names, column names, and sensitive information. 
There is no doubt that the WAFs (AWS, F5, and Cloudflare) will block them if we directly send them to the web service.
At this time, we input the above payloads into \mysystem and none of the three WAFs intercepted our carefully constructed payloads.
By analyzing the bypassing payloads, we can find that replacing whitespaces with control characters (\ie, \textit{Whitespace Substitution}) can bypass AWS. 
Furthermore, if we replace Data Manipulation Language (DML) tokens with inline comments, we can bypass F5 and Cloudflare.

In summary, these case studies further verify our conjecture: vendors' detection signatures are non-robust, causing severe vulnerabilities. 

\begin{figure}[htb!]
\centering
\includegraphics[width=0.49\textwidth]{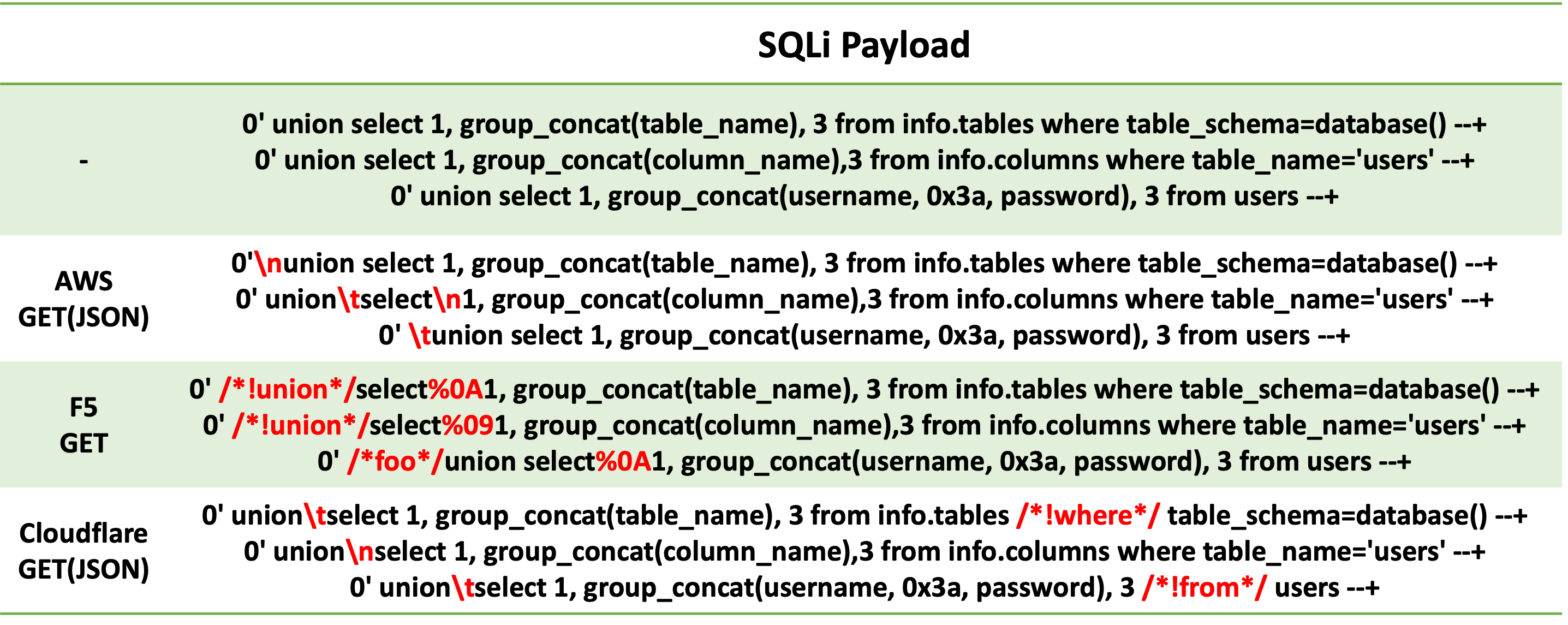}
\caption{Examples that can bypass WAF-as-a-service.}
\label{figure:examples-of-success-payloads-new}
\end{figure}

\begin{center}
\fcolorbox{black}{gray!5}{\parbox{0.99\linewidth}{
\textbf{Answer to RQ4: }Effective mutation methods for specific WAFs and different payloads are unique. Combining multiple mutation methods, \mysystem is much more effective in bypassing mainstream WAF-as-a-service solutions due to their vulnerable detection signatures for semantic matching and regular expression matching.}}
\end{center}

\section{Discussion}
\label{sec:6}
\subsection{Responsible Disclosure}
We reported the above results to the affected vendors by submitting vulnerability reports and contacting their technical support staff and security researchers via emails.
Up until now, three vendors have mitigated the flaws, and four vendors are working:

\textbf{F5}:
The F5 Security Incident Response Team responded to our email quickly and said that they would ``reach out to the respective team''.
Three weeks after our report, we were told that they had updated the rules for AWS WAF and the issue was addressed.

\textbf{Cloudflare}:
We submitted our findings to Cloudflare through its supporting system and the HackerOne platform.
After productive communications, we were informed that the WAF team had ``deployed some changes'', yet they would not tell us the details.

\textbf{Wallarm}:
They attached great importance to our report and declared that they ``are doing everything to resolve it as soon as possible''.
A month after our report, we learned from the update from the Wallarm detection team that the issue has been resolved and the ruleset for SQLi detection is now ``wider and more accurate''.

\textbf{AWS}:
The AWS Security Team replied to our email and proceeded to investigate it immediately.
After a thorough investigation, they said they ``will make a change to mitigate the behavior'' and will inform us ``once they've released the related improvements''.

\textbf{Fortinet} and \textbf{Cyber Security Cloud}:
After getting in touch with these vendors through their support email, we learned that they had confirmed the flaws and proceeded to solve them.
Their words are ``plan to address some of these'' and ``improve the rules in later update'', yet we have not received any new updates.

\textbf{ModSecurity}:
We reached the rules team (OWASP CRS) and engine team (Trustwave ModSecurity Security Team) respectively.
One CRS co-leader confirmed that CRS did have flaws and ModSecurity is insufficient in parsing query strings in JSON format, yet they can do little beyond digging through JSON with their rules.
Their developers are still working with us based on our detailed conclusions.
While the security researcher from ModSecurity said that the lack of JSON parsing is not due to some bug or malfunction.
ModSecurity would not parse JSON objects in query arguments or request headers by design.
Yet he admitted that ``the absence could sometimes make writing effective rules more cumbersome'', and they plan to implement the feature in ModSecurity v3.1.1.

\vspace{-2mm}
\subsection{Potential Defenses}
We conduct a preliminary exploration of two potential defense methods, \ie, adversarial training (AT) and pre-processing (PP).

\textbf{Adversarial training.}
We can improve the robustness of ML-based WAF through AT~\cite{goodfellow2014explaining_at}.
Firstly, we attack the $\text{CNN}^{1\%}$ model with 8852 SQLi payloads in the training set of HPD to obtain adversarial examples, using the \mysystemr attack method here for time considerations.
Next, we add the 1341 $\mathbf{x_{adv}}$ to the training set and retrain the $\text{CNN}^{1\%}$ model according to the previous settings.
Then, we re-attack the new model with SQLi payloads from the test set.
From Table~\ref{table:defense}, we can see that the performance of the CNN model dropped slightly after AT, \eg, the ASR of \mysystemr has dropped by 32.9\%, and the ASR of \mysystem has dropped by 30.59\%.
AT might be effective in defending \mysystem.
However, the limitation is that it needs to have sufficient adversarial samples for training or know the details of the attack strategy~\cite{li2018textbugger}.
Therefore, AT is limited in defending against unknown adversarial attacks because attackers usually do not disclose their methods. 

\textbf{Pre-processing.}
Just like dead code removal in programs, we use some pre-processing methods to remove the ``noise'' in SQLi payloads: \textit{unify the capitalization of letters}, \textit{remove comments}, \textit{remove control symbols}, and so on.
From Table~\ref{table:defense}, we can see that the ASRs dropped by less than 10\% on both attack modes.
Therefore, \mysystem is robust to common pre-processing methods.

\begin{table}[htb!]
\centering
\caption{The ASR results after defenses. AT means adversarial training and PP means pre-processing.}
\begin{tabular}{ccccc}
\toprule
\multirow{2}{*}{\begin{tabular}[c]{@{}c@{}}Defense\\ Mode\end{tabular}} & \multirow{2}{*}{\begin{tabular}[c]{@{}c@{}}AUC\\ (\%)\end{tabular}} & \multirow{2}{*}{\begin{tabular}[c]{@{}c@{}}Accuracy\\ (\%)\end{tabular}} & \multicolumn{2}{c}{ASR (\%)} \\
 &  &  & \mysystemr & \mysystemm \\
 \midrule
- & 99.959 & 99.90 & 7.94 & 26.46 \\
AT & 99.854 & 96.10 & 5.32 & 18.36 \\
PP & 99.959 & 99.90 & 6.32 & 25.18 \\
\bottomrule
\end{tabular}
\label{table:defense}

\end{table}

\subsection{Rethinking WAF-as-a-service}
Most real-world mainstream vendors employ the signature-based detection strategy combining semantic analysis and regular expression matching, which we have proven seriously vulnerable.
For ML-based WAF, the direct detection ability in the laboratory environment is passable, however, the effect of defending against adversarial attacks is not gratifying.
Intriguingly, there are vendors, such as Wallarm~\cite{website_wallarm}, already using ML technology to supplement the detection signatures adaptively and achieve feasible defense effects with low FNR and ASR.
This inspires us: can vendors build more promising WAFs by better utilizing AI methods or even implementing a WAF with multi-modal detection mechanisms?

\subsection{Limitations and Future Work}
There are some naturally inherent limitations to this kind of work.
Even though these do not affect the overall conclusion, we discuss them in the following for completeness.

\textbf{1) Dataset:}
We try our best to find open-sourced datasets, as there is no benchmark dataset for SQLi. 
One may argue that the baseline method~\cite{demetrio2020waf_wafamole} presented a dataset~\cite{website_wafamoledataset} containing both benign and malicious SQL samples.
However, in our task, benign samples should be ordinary HTTP request parameters, and malicious samples should be constructed HTTP request parameters containing SQL statements. 
Therefore, we have to exclude the dataset in~\cite{demetrio2020waf_wafamole} and only find two commonly used datasets of SIK and HPD probably due to the potential security risks or intellectual property.

\textbf{2) Verification of semantic-preserving:}
To the best of our knowledge, there is no practical way to check whether two SQLi payloads are semantic-equivalent other than comparing the running results~\cite{zhong2020squirrel}.
Therefore preliminary evaluations (RQ1) without the protection of WAF-as-a-service are performed.
Although the results show that our atomic-level mutation methods are semantic-preserving, we cannot guarantee strict semantic equivalence in all possible cases due to the assorted types of SQLi payloads.

\textbf{3) Manual definition of CFG:}
The manual definitions of CFG are to ensure that \mysystem will not invalidate the SQLi payloads or cause unexpected damage.
It may have limitations, such as making the generated payloads limited by prior knowledge.
Yet we have mitigated this by, \eg, recursively definition of CFG to enrich the diversity of the generated payloads.
Besides, as we have provided the corresponding interfaces, subsequent researchers who want to extend \mysystem only need to add simple entries in CFG.

Next, we will continue tracking vendors' feedback until they fix the flaws completely.
We shall implement the method of automatically analyzing and summarising the payloads based on the hierarchical tree to discover more generalized bypass patterns, which we could embed into other tools (\eg, sqlmap) or provide to vendors.
In addition, we plan to extend our framework to other domains, such as cross-site scripting attacks, webshell, \etc.
Besides, we will explore the integration of multi-factor authentication schemes into WAF systems, such as three-factor authentication based on extended chaotic maps and quantum-resistant two-factor authentication~\cite{review4_wang2021quantum2fa,review5_roy2018provably,review6_qiu2020practical}, to augment the defensive strength and robustness of web systems reliant on WAFs.
\section{Conclusion}
\label{sec:7}
We conduct the first systematic and comprehensive study on the security vulnerabilities of WAFs in the cloud, by proposing and implementing \mysystem.
With this general and extendable attack framework that works out of the box, attackers can bypass WAF-as-a-service solutions of different vendors for unique payloads effortlessly.
Extensive evaluation results demonstrate that \mysystem can effectively and efficiently bypass both state-of-the-art ML-based SQLi detectors and commercial WAF-as-a-service solutions from mainstream vendors.
Moreover, we condense out several fundamental deficiencies of real-world WAF-as-a-service, \eg, the vulnerable detection mechanisms, the non-robust signatures, and the flaws in parsing JSON-type parameters, which have successfully helped mainstream vendors improve their products.

\ifCLASSOPTIONcaptionsoff
  \newpage
\fi

\bibliographystyle{IEEEtran}
\bibliography{ref}

\appendix

\section{Context-Free Grammar}
\noindent \textbf{Context-Free Grammar}
\label{sec:appendix-cfg}
\begin{figure}[htb!]
    \centering
    \begin{lstlisting}[style=customstyle]
    $S_{\tau} \rightarrow S_{True} \text{ }|\text{ }   \tau_{string} \text{ }|\text{ }  \tau_{complex} \text{ }|\text{ }  \tau_{number} $
    $S_{\forall} \rightarrow f_{inline\_comment} \left( \forall \right) $
    $S_{\forall word} \rightarrow f_{swap\_case} \left( \forall_{word} \right) $
    $S_{\forall number} \rightarrow f_{change\_base} \left(\forall_{number} \right) $
    $S_{Where} \rightarrow \Sigma_{where}\ \cdot \ S_{False}\ \cdot \ \Sigma_{or} \text{ }|\text{ }  \Sigma_{where}\ \cdot \ S_{True}\ \cdot \ \Sigma_{and} $
    $S_{\vartextvisiblespace} \rightarrow \Sigma_{\vartextvisiblespace} \text{ }|\text{ }  S_{\gamma}$
    $S_{\gamma} \rightarrow S_{\gamma_{left}}\ \cdot \ S_{\gamma_{body}}\ \cdot \ S_{\gamma_{right}} $
    $S_{False} \rightarrow \Sigma_{false} \text{ }|\text{ }  S_{False}\ \cdot \ S_{\vartextvisiblespace}\ \cdot \  \Sigma_{or}\ \cdot \ S_{\vartextvisiblespace}\ \cdot \ \Sigma_{false} $
    $S_{True} \rightarrow \Sigma_{true} \text{ }|\text{ }  S_{True}\ \cdot \ S_{\vartextvisiblespace}\ \cdot \ \Sigma_{and}\ \cdot \ S_{\vartextvisiblespace}\ \cdot \ \Sigma_{true} $
    $S_{\gamma_{left}} \rightarrow \Sigma_{slash}\ \cdot \ S_{\gamma_{asterisk}}$
    $S_{\gamma_{right}} \rightarrow S_{\gamma_{asterisk}}\ \cdot \  \Sigma_{slash}$
    $S_{\gamma_{asterisk}} \rightarrow S_{\gamma_{asterisk}}\ \cdot \ \Sigma_{asterisk} \text{ }|\text{ }  \Sigma_{asterisk}$
    $S_{\gamma_{body}} \rightarrow \gamma_{chars} \ \cdot \  \gamma_{sentence} \ \cdot \ \gamma_{benign}$\end{lstlisting}%
    \caption{A simplified version of context-free grammar for generating semantic replacements. $S$ is the starting symbol; $\Sigma$, $\tau$, $f$ and $\gamma$ are all terminal symbols; Specifically, $f$ indicates that it needs to pass in values, such as changing the case of any word (line 3); $\tau$ means tautology and $\gamma$ refers to comments.}
    \label{figure:context-free-grammar}%
\end{figure}%

\begin{figure*}[htb!]
\centering
\includegraphics[width=0.79\textwidth]{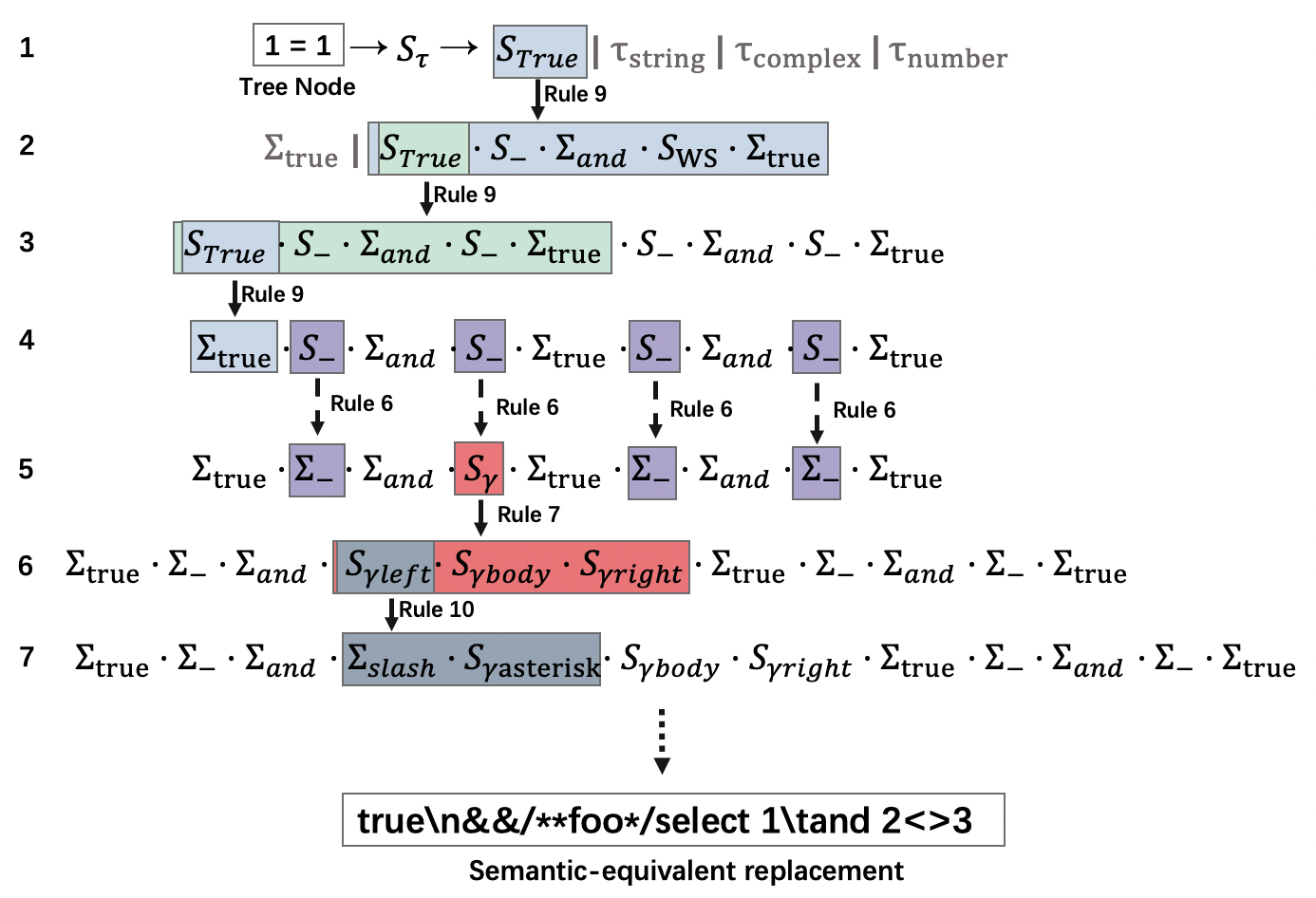}
\caption{An example of using context-free grammar to generate an equivalent replacement for ``1 = 1''. The contents of the same color connected by arrows indicate the states before and after one step generation}
\label{figure:cfg-demo}
\vspace{-5mm}
\end{figure*}

Here we describe the workflow of context-free grammar in \mysystem with an detailed example:
Suppose we have a node of ``1 = 1'' in the hierarchical SQLi representation tree.
According to its semantic of tautology, we can mutate it to terminal symbols of string-type tautology($\tau_{string}$, like ```foo' like `foo''' in Table~\ref{table:examples-of-terminal-characters}), complex tautology ($\tau_{complex}$, such as ``(select ord('r') regexp 114) = 0x1''), number-type tautology ($\tau_{number}$, such as ``0x2 = 2'') or feed it into an entry of ``true'' boolean expression ($S_{True}$) for further processing, as shown in the first line of Figure~\ref{figure:context-free-grammar}.
Here the algorithm chooses the latter one, \ie, line 1$\rightarrow$line 2 in Figure~\ref{figure:cfg-demo}.
Deep into rule 9 of Figure~\ref{figure:context-free-grammar}, $S_{True}$ can be mutated to terminal symbols of ``true'' boolean expression (\eg, 1, select 1, 2$<>$3 in Table~\ref{table:examples-of-terminal-characters}) or be mutated into a recursive nested form combined with whitespaces and ``or'' expression ($S_{True}\ \cdot S_{\vartextvisiblespace}\ \cdot \Sigma_{and}\ \cdot S_{\vartextvisiblespace}\ \cdot\ \Sigma_{true}$).
Afterwards, we can see that it uses rule 9, rule 6, rule 7 and rule 10 in Figure~\ref{figure:context-free-grammar} respectively, and finally generates a candidate SQLi payload ``true\textbackslash{}n\&\&/**foo*/select 1\textbackslash{}tand 2$<>$3'', which is semantically equivalent to ``1 = 1''.

It is paramount that the context-free grammar (including starting symbols, non-terminal symbols, terminal symbols, and predefined rules) are manually defined, which is to ensure that \mysystem will not invalidate the generated SQLi payloads or cause unexpected damage.
Subsequent researchers only need to add simple items in the interfaces of our grammar if they want to extend it.

\begin{table}[htbp]
\centering
\caption{Examples of terminal symbols in CFG. $\Sigma$, $\tau$, $f$ and $\gamma$ are all terminal symbols; Specifically, $f$ indicates that it needs to pass in values, such as changing the case of any word (line 3 of Figure 1); $\tau$ means tautology and $\gamma$ refers to comments.} 
\begin{tabular}{@{}ll@{}}
\toprule
$\Sigma$ & Examples \\ \midrule
$\Sigma_{slash}$ & / \\
$\Sigma_{asterisk}$ & * \\
$\Sigma_{\vartextvisiblespace}$ & \textvisiblespace ,\textbackslash{}t, \textbackslash{}n \\
$\Sigma_{or}$ & or, $||$ \\
$\Sigma_{and}$ & and, \&\& \\
$\Sigma_{false}$ & 0, select 0, 2=3, false \\
$\Sigma_{true}$ & 1, select 1, 2\textless{}\textgreater{}3, true \\
$\gamma_{chars}$ & a21r*!, skx9\{1-, 2sd.\$5s, sd2j3znm, 39nsq1 \\
$\gamma_{sentence}$ & hello world, today is dog, over the alphabet \\
$\gamma_{benign}$ & provincia=burgos, login=karina9, pic=1.gif \\
$f_{inline\_comment}$ & union $\rightarrow$ /*!union*/, where $\rightarrow$ /*!where*/ \\
$f_{swap\_case}$ & select $\rightarrow$ sElECt, OR $\rightarrow$ or, from $\rightarrow$ FrOM \\
$f_{change\_base}$ & 1 $\rightarrow$ 0x1, 256 $\rightarrow$ 0x100, 176 $\rightarrow$ select 176 \\
$\tau_{number}$ & 78 = 78, 1 = 1, 2021 like 2021, 0x2 = 2 \\
$\tau_{string}$ & `6' like `6', `foo' like `foo', `bar' = `bar' \\
$\tau_{complex}$ & \begin{tabular}[c]{@{}l@{}} {(select ord(`r') regexp 114) = 0x1,} \\ {(select 1) = (select ord(`r') between 114 and 115)}\end{tabular} \\
\bottomrule
\end{tabular}
\label{table:examples-of-terminal-characters} 
\end{table}

In addition to line 9 in Figure~\ref{figure:context-free-grammar}, some of the other rules (\eg, line 8 and line 12) are also defined recursively, which might cause an endless loop, \ie, terminal symbols can never be reached in generating the candidate SQLi payload.
To address this issue, instead of simply defining the max depth of recursion, we present a weighted mutation strategy (Algorithm~\ref{algo:cfg}) by employing a decay rate $D$ (\eg, 0.5).
The weight of a non-terminal character $V$ being selected will decrease exponentially according to $D$, \eg, the weight of selecting $V$ for the third time is $0.5^{3}$.

\begin{algorithm}[htb!]
\caption{Weighted CFG Generation}
\label{algo:cfg}
\SetKwFunction{WeightedChoice}{WeightedChoice}
\SetKwFunction{Self}{Self}
\KwData{CFG Rules $R$, Entry Symbol $sym$, Decay Rate $D$, Chosen Counts $C$}
\KwResult{Generated sentence $sentence$ }
$sentence \leftarrow EmptyString $ \;
$weights \leftarrow EmptyList $ \;
\For{e in R[sym]}{\If{e in C}
{weights.append$(D^{C[e]})$}
\Else{weights.append$(1.0)$}
}
\tcp{Weighted selection function.} 
$rand\_sym \leftarrow$ $R$[$sym$][\WeightedChoice$(weights)$]   \;
$C$[$rand\_sym$] $\leftarrow C$[$rand\_sym$]$ + 1$ \;
\For{s in rand\_sym}{
\If{s in R}{
\tcp{Call itself recursively.} 
$sentence \leftarrow sentence + $\Self($R, s, D, C$) \;
}
\Else{
$sentence \leftarrow sentence + s$ \;
}
}
$C$[$rand\_sym$] $\leftarrow C$[$rand\_sym$]$ - 1$ \;
return $sentence$ \;
\end{algorithm}

Besides, it is worth mentioning that the above-mentioned context-free grammar (including starting symbols, non-terminal symbols, terminal symbols, and predefined rules) are manually defined, which is to ensure that \mysystem will not invalidate the generated SQLi payloads or cause unexpected damage.
Subsequent researchers only need to add simple items in the interfaces of our grammar if they want to extend it.


\section{Experimental performance consumption}
\noindent \textbf{Experimental performance consumption}

As the findings in RQ2, \mysystem demonstrates a capability to achieve a higher attack success rate with fewer interaction steps.
To provide a deeper understanding of the associated time and performance overheads, we conduct the following experiments.

We reimplement the experiments in RQ2, which are carried out on an Ubuntu 20.04 server, equipped with 64GB RAM and powered by an Intel(R) Xeon(R) Silver 4210R CPU.
The primary objective was to assess both the runtime and memory consumption.
It's important to emphasize that the reported memory consumption values represent the average taken from continuous sampling throughout the entire experiment duration. 
We benchmarked against two baseline methods, \mole and \rl, and also assessed both AdvSQLi(A) – a simplified version of our primary approach and AdvSQLi, our core method.

As shown in Table~\ref{table:attack-consumption}, the memory consumption across different methods is fairly consistent, with only minor differences observed.
In most scenarios, \mysystem consumes about 10\% additional memory compared to the other techniques, marking its competitive efficiency.
\rl emerges as the fastest method.
This speed can be attributed to its pre-completed training phase.
During the attack, \rl only needs to select the most suitable mutation method based on the current payload state.
As anticipated, \mysystem does demand the most time, which is closely tied to the construction and search processes of its Monte Carlo tree.
Although \mysystem exhibits a higher time overhead, its vastly superior attack success rate compared to the baseline methods justifies this cost.
When we distribute this time consumption across 1000 samples, the results provide a worthy trade-off for its efficacy.

\begin{table*}[htb]
\centering
\caption{\upshape The consumption results of attacking against \local under the settings same with \#Table V. \texttt{WAM} means \mole. The time and memory consumption of \rl only involves the inference phase and does not include the pre-training phase. The time consumption of attacking WAF-Brain models is limited by the inference speed.}
\begin{tabular}{cccccccccc}
\toprule
\multirow{3}[4]{*}{Dataset} & \multirow{3}[6]{*}{\tabincell{c}{Target SQLi\\Detection}} &   \multicolumn{8}{c}{Consumption}\\
\cmidrule{3-10}   &  & \multicolumn{4}{c}{Time (s)} & \multicolumn{4}{c}{Memory (MB)} \\
\cmidrule(r{1pt}){3-6} \cmidrule(l{1pt}){7-10}    &     & \mysystemr & \mysystem & \texttt{WAM} & \rl & \mysystemr & \mysystem & \texttt{WAM} & \rl    \\
\midrule
\multirow{7}{*}{HPD} & $\text{WAF-Brain}^{\delta=0.1}$  &  \textbf{554.53} & 2240.03 & 4713.73 & 1751.27 & \textbf{344.55} & 369.34 & 355.98 & 397.46 \\
 & $\text{WAF-Brain}^{1\%}$  &  \textbf{135.02} & 315.48 & 577.12 & 564.39 & \textbf{344.08} & 365.68 & 361.01 & 388.27  \\
 & $\text{WAF-Brain}^{1\perthousand}$  &  \textbf{18.42} & 51.97 & 198.96 & 53.76 & \textbf{342.88} & 362.49 & 343.59 & 382.28  \\
 & $\text{CNN}^{1\%}$  & 293.44 & 1161.01 & 1029.56 & \textbf{198.52} & 350.72 & 370.13 & \textbf{340.38} & 376.7
 \\
 & $\text{CNN}^{1\perthousand}$   & 290.83 & 982.02 & 950.73 & \textbf{179.91} & 351.07 & 370.53 & \textbf{343.74} & 373.76 \\
 & $\text{LSTM}^{1\%}$ & 654.28 & 2218.46 & 1036.59 & \textbf{223.82} & \textbf{350.55} & 371.04 & 355.3 & 375.66 \\
 & $\text{LSTM}^{1\perthousand}$ & 639.42 & 2154.88 & 983.12 & \textbf{221.69} & \textbf{350.41} & 370.82 & 350.55 & 383.88 \\
 \midrule
\multirow{7}{*}{SIK} & $\text{WAF-Brain}^{\delta=0.1}$  &  \textbf{483.96} & 1817.85 & 5434.94 & 1342 & \textbf{346.52} & 372.88 & 356.02 & 373.04 \\
 & $\text{WAF-Brain}^{1\%}$   & \textbf{91.95} & 183.19 & 362.42 & 383.37 & \textbf{339.86} & 367.58 & 343.88 & 379.64  \\
 & $\text{WAF-Brain}^{1\perthousand}$ &  \textbf{6.82} & 21.51 & 52.23 & 26.8 & \textbf{341.22} & 364.02 & 356.68 & 366.52  \\
 & $\text{CNN}^{1\%}$ & 266.19 & 1015.42 & 1009.42 & \textbf{183.85} & \textbf{350.77} & 376.21 & 355.51 & 378.28 \\
 & $\text{CNN}^{1\perthousand}$ & 252.28 & 976.15 & 955.46 & \textbf{179.08} & \textbf{351.45} & 378.1 & 354.65 & 370.32 \\
 & $\text{LSTM}^{1\%}$ &  525.8 & 1925.24 & 954.07 & \textbf{232.8} & \textbf{352.42} & 377.26 & 357.17 & 373.65 \\
 & $\text{LSTM}^{1\perthousand}$ &  509.53 & 1865.95 & 920.89 & \textbf{223.93} & \textbf{349.9} & 376.39 & 356.19 & 375.16 \\
\bottomrule
\end{tabular}
\label{table:attack-consumption}
\end{table*}

\begin{IEEEbiography}[{\includegraphics[width=1in,height=1.25in,clip,keepaspectratio]{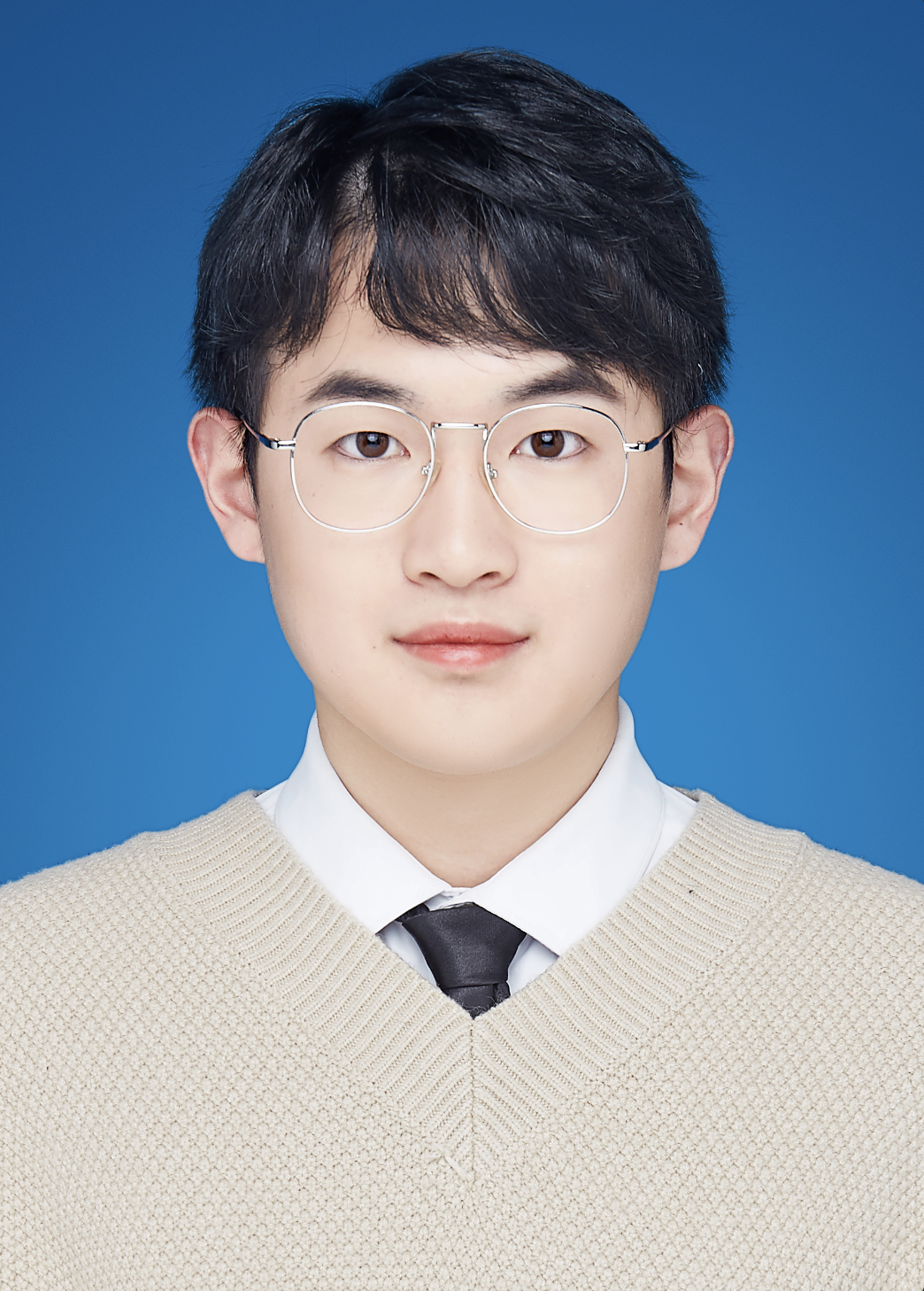}}]{Zhenqing Qu}
is a graduate student at Zhejiang University. His research focuses on web security and data-driven security. He has spoken at Black Hat Asia 2022. He is a member of the Azure Assassin Alliance CTF Team. Additionally, he has reported several severe defects to mainstream security vendors, which were confirmed and fixed quickly.
\end{IEEEbiography}

\vspace{-10mm}

\begin{IEEEbiography}[{\includegraphics[width=1in,height=1.25in,clip,keepaspectratio]{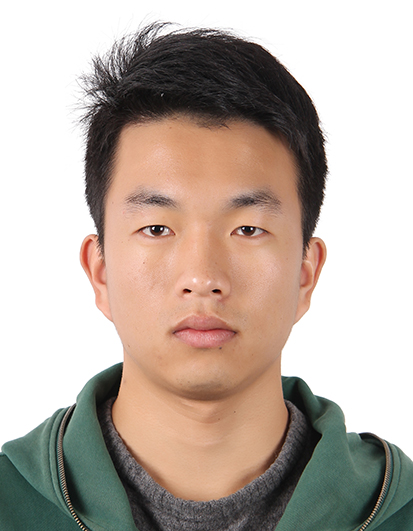}}]{Xiang Ling}
is currently an assistant professor at Institute of Software, Chinese Academy of Sciences. He received his PhD degree from Zhejiang University. His research focuses on data-driven security, AI security and program analysis.
His work has been published top-ranked conferences and journals, like IEEE S\&P, INFOCOM, ICSE, TNNLS and TKDD.
\end{IEEEbiography}

\vspace{-10mm}

\begin{IEEEbiography}[{\includegraphics[width=1in,height=1.25in,clip,keepaspectratio]{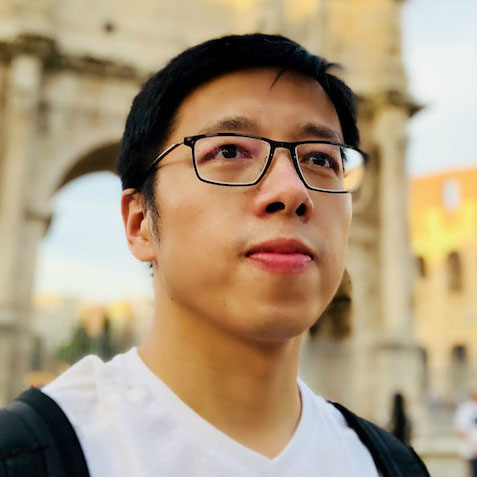}}]{Ting Wang}
is an Associate Professor and Empire Innovation Scholar in the Department of Computer Science at Stony Brook University. He received his Ph.D. degree from Georgia Tech. He conducts research at the interface of machine learning, privacy, and security. His recent work focuses on improving AI technologies in terms of security assurance, privacy preservation, and decision-making transparency.
\end{IEEEbiography}

\vspace{-10mm}

\begin{IEEEbiography}[{\includegraphics[width=1in,height=1.25in,clip,keepaspectratio]{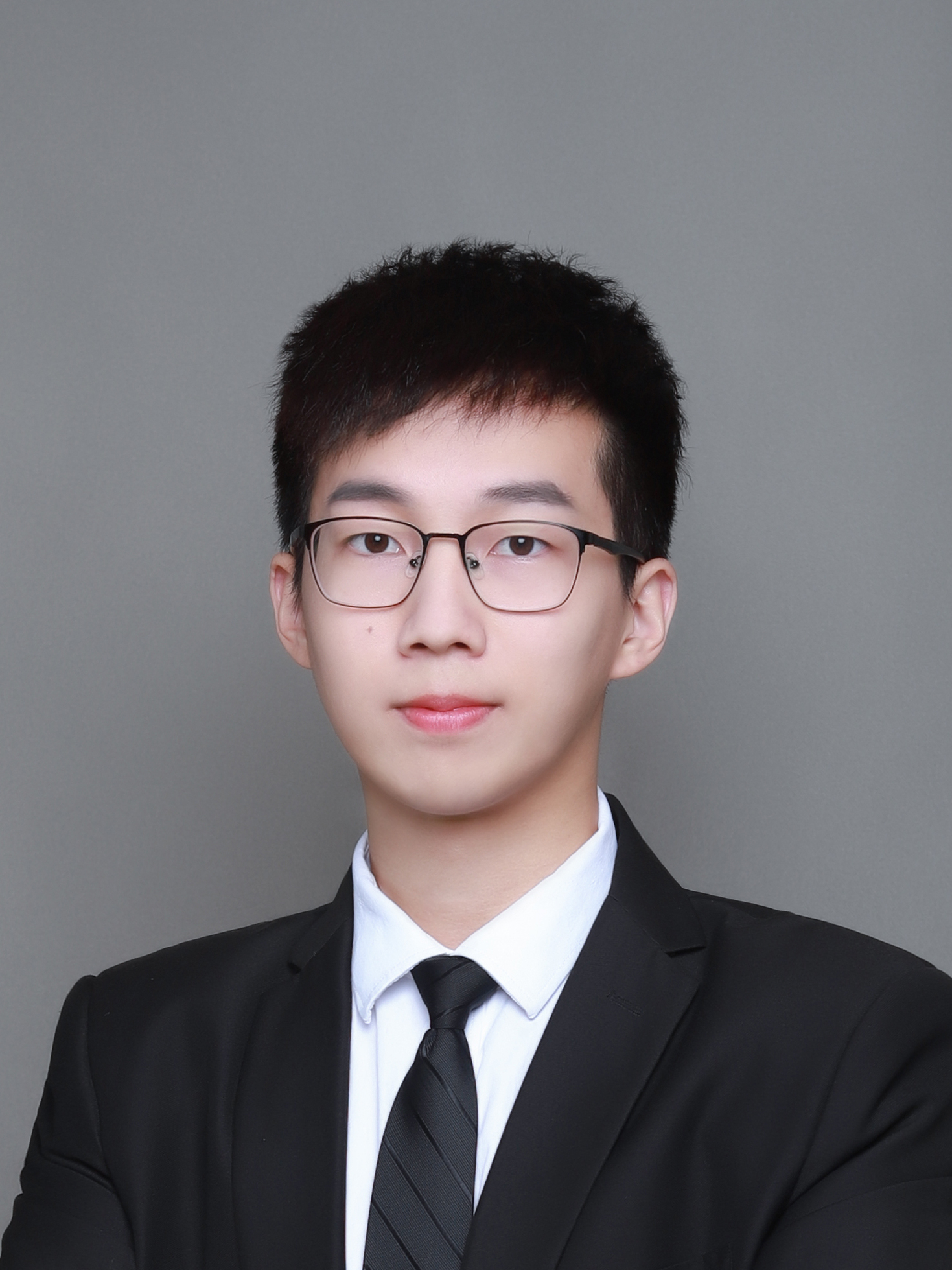}}]{Xiang Chen}
received the B.Eng. and M.Eng. degrees from Fuzhou University, in 2019 and 2022, respectively. He is currently pursuing the Ph.D. degree with the College of Computer Science and Technology, Zhejiang University, China. He has published papers in IEEE INFOCOM and IEEE ICNP. He received the Best Paper Award from IEEE/ACM IWQoS 2021 and the Best Paper Candidate from IEEE INFOCOM 2021. His research interests include programmable networks and network security.
\end{IEEEbiography}

\vspace{-10mm}

\begin{IEEEbiography}[{\includegraphics[width=1in,height=1.25in,clip,keepaspectratio]{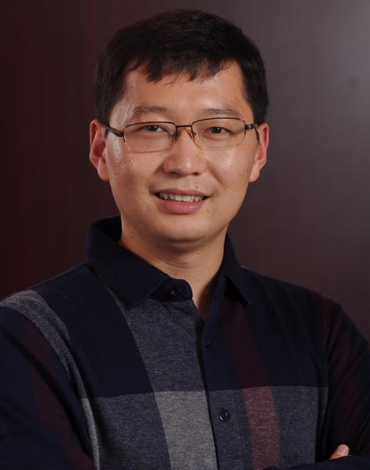}}]{Shouling Ji}
is a ZJU 100-Young Professor in the College of Computer Science and Technology at Zhejiang University and a Research Faculty in the School of Electrical and Computer Engineering at Georgia Institute of Technology (Georgia Tech). He received a Ph.D. degree in Electrical and Computer Engineering from Georgia Institute of Technology , a Ph.D. degree in Computer Science from Georgia State University, and B.S. (with Honors) and M.S. degrees both in Computer Science from Heilongjiang University. His current research interests include Data-driven Security and Privacy, AI Security and Big Data Analytics. He is a member of ACM, IEEE, and CCF and was the Membership Chair of the IEEE Student Branch at Georgia State University (2012-2013).
\end{IEEEbiography}

\vspace{-10mm}

\begin{IEEEbiography}[{\includegraphics[width=1in,height=1.25in,clip,keepaspectratio]{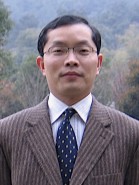}}]{Chunming Wu}
is currently a Professor with the College of Computer Science and Technology, Zhejiang University. He is also the Associate Director of the Research Institute of Computer System Architecture and Network Security, Zhejiang University, and the Director of the Fluctlight Security Lab. His research interests include network security, reconfigurable network and next-generation network infrastructures. He has published more than 90 papers in a series of international journals, magazines, and conferences, e.g., CCS, S\&P, USENIX, INFOCOM, ToN, etc.
\end{IEEEbiography}

\end{document}